%% file: sensys27.tex
\documentclass[sigplan,table,svgnames,nonacm]{acmart}

\usepackage[utf8]{inputenc}
\usepackage{amsmath}
\usepackage{graphicx}
\usepackage{hyperref}
\usepackage{booktabs} %
\usepackage{enumitem} %
\usepackage{xcolor} %
\usepackage{fontawesome5} %
\graphicspath{{assets/figures/}}
\usepackage{url} %

\input{macros}

\usepackage{array}
\usepackage{makecell}

\usepackage{multirow}

\usepackage[most]{tcolorbox}
\usepackage{tabularx}

\definecolor{calmblue}{RGB}{0, 56, 168}
\definecolor{modblue}{RGB}{102, 178, 255}
\definecolor{unstableyellow}{RGB}{255, 221, 51}
\definecolor{aggressiveorange}{RGB}{255, 153, 51}
\definecolor{aggressivered}{RGB}{255, 77, 77}
\definecolor{capgreen}{RGB}{46,125,50}
\definecolor{flashfusion}{RGB}{47,143,87}
\definecolor{capamber}{RGB}{191,144,0}
\definecolor{capred}{RGB}{198,40,40}

\newcommand{\cmark}{\textcolor{capgreen}{\checkmark}}
\newcommand{\pmark}{\textcolor{capamber}{\(\sim\)}}
\newcommand{\xmark}{\textcolor{capred}{\(\times\)}}

\AtBeginDocument{%
}

\setcopyright{acmlicensed}
\copyrightyear{2026}
\acmYear{2026}
\acmDOI{10.1145/0000000.0000000}

\acmConference[SenSys]{ACM/IEEE International Conference on
Embedded Artificial Intelligence and Sensing Systems}{May 17--20, 2027}{Boulder, CO, USA}
\acmBooktitle{ACM/IEEE International Conference on
Embedded Artificial Intelligence and Sensing Systems, May 17--20, 2027, Boulder, CO, USA}
\acmISBN{978-1-4503-0000-0/25/05}

\begin{document}

\title{Flash-Fusion: Fast, Expressive Queries over Sensor Streams with LLMs}

\author{Kausar Patherya}
\affiliation{%
\institution{Georgia Institute of Technology}
\city{Atlanta}
\state{Georgia}
\country{USA}}
\email{kpatherya3@gatech.edu}

\author{Ashutosh Dhekne}
\affiliation{%
\institution{Georgia Institute of Technology}
\city{Atlanta}
\state{Georgia}
\country{USA}}
\email{dhekne@gatech.edu}

\author{Francisco Romero}
\affiliation{%
\institution{Georgia Institute of Technology}
\city{Atlanta}
\state{Georgia}
\country{USA}}
\email{faromero@gatech.edu}

\input{tex/abstract.tex}


\maketitle

\pagestyle{standardpagestyle}

\input{tex/introduction.tex}

\input{tex/challenges.tex}

\input{tex/design.tex}

\input{tex/implementation.tex}

\input{tex/evaluation.tex}

\input{tex/relatedwork.tex}

\input{tex/conclusion.tex}

\bibliographystyle{ACM-Format}
\bibliography{refs.bib}

\end{document}

%% file: macros.tex
\setlist[itemize]{leftmargin=*}

\newcommand*{\eg}{\emph{e.g.,}~}

\newcommand{\minihead}[1]{{\vspace{.45em}\noindent\textbf{#1.~~}}}

\usepackage{etoolbox} %
\newtoggle{showmarks}
 \toggletrue{showmarks}

\iftoggle{showmarks}{
  
  \newcommand{\ashutosh}[1]{\textcolor{brown}{AD: #1}}
  \newcommand\franky[1]{\textcolor{olive}{FR: #1}}
  \newcommand\kausar[1]{\textcolor{teal}{KP: #1}}
  \newcommand\update[1]{\textcolor{blue}{* #1}}
  \newcommand\auto[1]{\textcolor{gray}{CP: #1}}
}{
  
  \newcommand{\ashutosh}[1]{\unskip}
  \newcommand\franky[1]{\unskip}
  \newcommand\kausar[1]{\unskip}
  \newcommand\update[1]{\unskip}
  \newcommand\auto[1]{\unskip}
}

%% file: tex/abstract.tex
\begin{abstract}

Pervasive sensor deployments have generated significant interest in IoT data analysis across fields such as healthcare, transportation and urban planning.
At the same time, the rise of Large Language Models (LLMs) offer a new tool for exploring IoT data through their expressive, natural language interface.
Unfortunately, natural language querying of IoT sensor streams is impractical today: raw telemetry exhausts LLM context windows in fractions of a second at scale, incurs prohibitive per-token costs, and produces hallucinated responses.
Existing systems work around these limits with lossy text summarization or unconstrained code synthesis, producing expensive, unverified computations.
Multi-step queries compound this problem: logical errors in intermediate operations can derail the final answer.

We present Flash-Fusion, an agentic framework for natural-language querying of tabular IoT data.
Flash-Fusion identifies which available data fields and supported derivations can answer the user's query, constructs a plan using typed, domain-specific operators, and validates its structure and schema compatibility before executing.
Because constructing plans still requires costly LLM reasoning, Flash-Fusion caches validated operator sequences for reuse across compatible queries.
A smaller LLM adapts a cached plan's parameters to each query, cutting latency and cost.

We evaluate Flash-Fusion across human activity recognition, ECG monitoring, and vehicular telematics, covering direct lookups, multi-step reasoning, predictive tasks, and out-of-scope requests.
Flash-Fusion achieves 92.2\% query accuracy versus 76.1\% for ReAct, the strongest baseline, while completing queries \(7\times\) faster and consuming \(5\times\) fewer tokens.
Flash-Fusion enables users across disciplines to explore IoT data without manually constructing queries.

\end{abstract}

%% file: tex/introduction.tex
\section{Introduction}
\label{sec:introduction}

\begin{figure}[t]
  \centering
  \includegraphics[width=\linewidth]{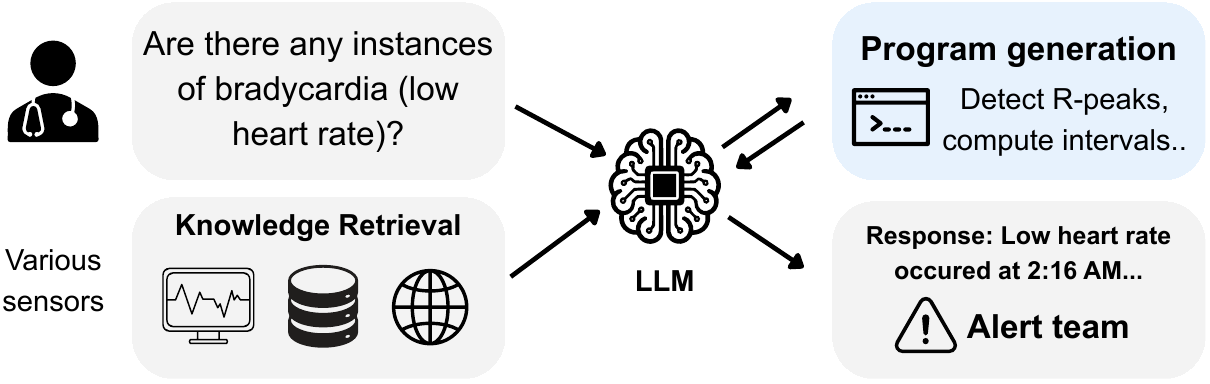}
  \caption{
    \textbf{LLMs and sensor data.}
    A medical technician asks an LLM a natural-language question about ECG data.
  }
  \label{fig:overview}
\end{figure}

The Internet of Things (IoT) generates continuous, high-volume streams of sensor data across domains such as healthcare, transportation, and industrial monitoring~\cite{houssein2024internet}.
Large Language Models (LLMs) increasingly process this data directly~\cite{zong2025integrating, an2024iot}, benefiting from stronger reasoning~\cite{xu2025towards} and larger context windows~\cite{liu2025comprehensive}.
LLMs enable users to query sensor data in plain English (NL2SQL), operating as a natural language layer over existing ETL (Extract, Transform, Load) pipelines~\cite{mussah2023accelerating, li2024dawn}.
For example, a medical technician can directly ask, \textit{``Are there any instances of bradycardia (low heart rate) in patient 101?''}~\cite{al2025vega, apandi2018arrhythmia}, instead of manually writing a SQL query (Figure~\ref{fig:overview}).

\begin{figure}[t]
  \centering
  \includegraphics[width=1.0\linewidth]{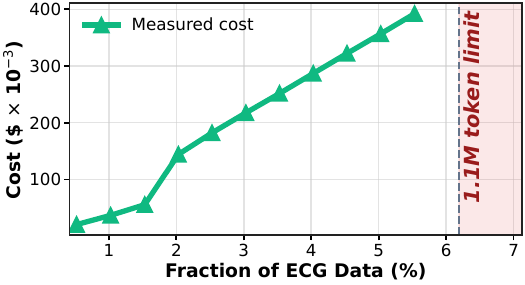}
  \caption{\textbf{Cost bottlenecks with directly prompting an LLM with sensor data.}
  Input-token cost grows with the amount of daily ECG data placed in the prompt; only $\sim$6\% fits within the tested LLM's 1.1M-token limit.}
  \label{fig:ministudy}
\end{figure}

The amount of sensor data is large and continues to grow, making direct LLM processing impractical. An ECG machine operating at 200\,Hz produces $\sim 10^{7}$ measurements per patient per day~\cite{couderc2012telemetric}.
This volume creates three challenges.
First, serializing the measurements into text exceeds the LLM's context window~\cite{stabler2025impact}.
Second, processing larger inputs increases latency and cost even before the context limit is reached.
As Figure~\ref{fig:ministudy} shows, only $\sim$6\% of a day's ECG measurements fit within the tested LLM's 1.1M-token context window.
Streaming the data in batches avoids this limit but still pays for every input token.
In this hospital setting, we estimate monthly input-token costs of $\sim$\$938K~\cite{comanici2025gemini}.
Third, the medical technician must determine whether the answer generated by the LLM uses the intended ECG data and computes the requested quantity.
Checking these decisions against high-volume telemetry demands time and domain expertise.
Practical LLM-based querying therefore requires limiting the data placed in the prompt and verifying that the computation matches the user's request.

Existing approaches address different aspects of these challenges.
Direct-prompting approaches, such as HARGPT~\cite{ji2024hargpt}, remain constrained by the LLM's context window (Figure~\ref{fig:ministudy}). LLMSense~\cite{ouyang2024llmsense} reduces input size through natural-language summarization, a common way to extend effective context~\cite{fei2024extending}.
However, summaries may omit readings needed for later queries, ruling out exact aggregations or threshold comparisons on values the summary did not keep.

Program-synthesis approaches, such as AutoIOT~\cite{shen2025autoiot}, and tool-based approaches, such as ReAct~\cite{yao2022react}, move computation outside the prompt by generating code or invoking tools over the underlying data.
For the simple ECG query, selecting the relevant fields, filtering readings to patient 101, and checking for low heart rate requires only a short sequence of operations.
The difficulty grows for multi-hop queries: an incorrect intermediate filter, grouping, or aggregation can propagate through subsequent steps and produce a plausible but incorrect answer.
Both approaches reason before acting: AutoIOT drafts an algorithm and refines its code with execution feedback, and ReAct interleaves reasoning, actions, and observations.
Yet code that executes successfully can still select the wrong data or apply the wrong transformations.
Producing the correct answer therefore depends on the planner's ability to translate the user's high-level goal (\eg locate instances of low heart rate) into the appropriate data fields and operations.
This motivates a framework that supports this translation through explicit operator plans and checks their validity before execution.

We present Flash-Fusion, an agentic framework for querying tabular IoT sensor data (Figure~\ref{fig:overview}).
Given a user query, its planner identifies which data fields and supported derivations can answer the question, then composes the operators needed to compute the result.
Plan validators then check operator names, required parameters, parameter types, and referenced columns.
The runtime executes predefined operator implementations, not LLM-generated code.

Constructing a plan requires an LLM call to select and order operators, accounting for over 70\% of end-to-end latency on average.
Providing the planner with descriptions of the available operators also increases input-token consumption and cost.
We observe that some queries can be answered using the same operator sequence, changing only parameter values such as column references and filter thresholds.
We therefore design and implement a plan cache that stores previously validated operator sequences.
When the cache retrieves a candidate plan, a smaller LLM sets its column references and other parameters for the new query rather than constructing the operator sequence anew.
Flash-Fusion validates the adapted plan before execution and invokes the full planner when no candidate is retrieved or errors arise in parameter grounding.
By reusing operator sequences and adapting only their parameters, the cache reduces average query cost by 2.1$\times$ compared with planning every query from scratch (Section~\ref{sec:ablations}).

Flash-Fusion enables personas across disciplines - safety officers, urban planners, fleet managers, and data scientists - to quickly and efficiently iterate over IoT data without the burden of manually tuning, refining, and verifying queries.
We evaluate Flash-Fusion on three real-world IoT datasets spanning human activity recognition~\cite{weiss2019wisdm}, medical informatics~\cite{apandi2018arrhythmia}, and fleet management~\cite{killeen2019iot}, using a diverse query benchmark covering direct lookups, multi-step reasoning, predictive tasks, and out-of-scope requests.
Averaged across datasets, Flash-Fusion is more accurate than the strongest baseline, ReAct~\cite{yao2022react} (92.2\% vs.\ 76.1\%), completing queries 7$\times$ faster and consuming 5$\times$ fewer tokens.

In summary, our contributions are the following:
\begin{enumerate}[leftmargin=*, itemsep=1pt]
    \item We characterize and identify challenges in using LLMs for large-scale IoT analysis.
    \item We design Flash-Fusion, which answers natural-language queries over IoT data by composing typed operators, validating plans before execution, and reusing them through a plan cache.
    \item We show that Flash-Fusion improves query accuracy, latency, and cost over state-of-the-art systems.
    \item As part of our evaluation, we collect vehicular transportation data from a live bus fleet for the community to use in their IoT analysis research.
\end{enumerate}

%% file: tex/challenges.tex
\section{Challenges and Limitations}
\label{sec:challenges}

Answering a natural-language query over sensor data requires the system to access the telemetry and compute the right answer from it.
Throughout, a sensor table has one row per reading and one column per sensor channel.
For example, a 200\,Hz ECG table collects a row every 5\,ms with columns for timestamp, lead I, and lead II.
We revisit the bradycardia query from Section~\ref{sec:introduction}.

Prior IoT--LLM systems follow three approaches, each of which fails on this query at a different point.
Direct prompting exceeds the context window and cost budget (R1).
Program synthesis leaves query-to-column mapping unchecked, so it cannot reliably ground or reject a request (R2).
Chain-of-thought guided synthesis struggles to compose multiple steps correctly (R3).
Table~\ref{tab:capability_matrix} summarizes how these systems fare; the subsections below explain each failure.

\begin{table}[t]
\centering
\caption{\textbf{Capabilities of prior IoT--LLM systems and Flash-Fusion.}
\cmark: supported; \pmark: partial support; \xmark: not supported.
}
\label{tab:capability_matrix}
\footnotesize
\setlength{\tabcolsep}{2.0pt}
\renewcommand{\arraystretch}{1.15}
\begin{tabular}{@{}lccccc@{}}
\toprule
\textbf{Capability} & \textbf{HARGPT} & \textbf{LLMSense} & \textbf{AutoIOT} & \textbf{ReAct} & \textbf{Ours} \\
\midrule
General-purpose        & \pmark & \cmark & \cmark & \cmark & \cmark \\
Scalable               & \xmark & \pmark & \cmark & \cmark & \cmark \\
Deterministic          & \xmark & \xmark & \xmark & \xmark & \cmark \\
Grounded to schema     & \xmark & \xmark & \xmark & \cmark & \cmark \\
Rejects out-of-scope   & \xmark & \xmark & \xmark & \pmark & \cmark \\
\bottomrule
\end{tabular}
\end{table}

\subsection{Context and cost scaling}
\label{sec:context}

Direct prompting gives an LLM access to sensor data by serializing observations as text and placing them beside the question.
HARGPT~\cite{ji2024hargpt} places raw readings in the prompt. LLMSense~\cite{ouyang2024llmsense} instead has the model narrate each chunk of readings and reasons over the narrations.

Raw prompting fails first as the context size grows.
At 200\,Hz, one patient produces $\sim 1.7\times10^{7}$ rows per day.
At roughly five tokens per serialized row (a timestamp and two lead values), a day of data is $\sim 86$M tokens, or 86$\times$ a 1M-token context window~\cite{gpt-55}.
Figure~\ref{fig:ministudy} measures the effect: the tested model (context window of 1.1M tokens) fits under 6\% of a day's measurements, so it misses the slow heart rate episodes that define bradycardia~\cite{hadiyoso2015arrhythmia}.
Splitting the data into chunks avoids the limit but still bills every token: a single query over one day of ECG telemetry costs over \$430~\cite{groqpricing}, or \$13K per month.

Summarized prompting reduces input tokens but changes what the model can answer.
A summary is useful only if it retains what the query needs, and it cannot recover a brief low heart rate episode it never recorded.
Like raw prompting, summarization cost grows with data size, since more data requires more summarization calls.

\vspace{0.5em}
\noindent \textbf{R1 (bounded context):}
The system must answer over the full table while sending the LLM only a small, fixed amount of data and preserving the exact values the query needs.

\subsection{Grounding and answerability}
\label{sec:hallucinations}

Program synthesis removes the scaling bottleneck of direct prompting but does not check whether the table can answer the query.
AutoIOT~\cite{shen2025autoiot} retrieves background knowledge, drafts an algorithm, generates code, and refines it using execution feedback.
Because the table is read by code, the prompt no longer grows with the number of rows, which addresses R1.
However, code that runs successfully does not show that the table contains the fields the query needs.
AutoIOT's retrieval informs which algorithm to use, not whether the dataset supports it.

This gap produces two failures. First, the request may need a field that does not exist.
To illustrate, consider the query \textit{``Predict this user's location.''} issued against the WISDM dataset~\cite{weiss2019wisdm}, which contains only accelerometer data.
Rather than rejecting the query, the LLM may train a classifier on the accelerometer axes and return a location that was never measured.
Language models tend to produce a fluent answer rather than abstain when uncertain~\cite{kalai2025language}.
In safety-critical domains such as remote health monitoring, such plausible yet hallucinated responses carry direct risk~\cite{roustan2025clinicians}.

Second, the code may use an existing field under the wrong meaning.
For example, code that treats lead voltage as heart rate still returns a number, but answers another question.

A column name alone does not decide answerability.
Our ECG table has no heart rate column, but heart rate is derivable by detecting R-peaks and converting their spacings to beats per minute, so our bradycardia query is answerable.
Location, on the other hand, is not derivable from accelerometer axes at all.
A system must distinguish a quantity derivable from existing fields from information absent from the data.

\vspace{0.5em}
\noindent \textbf{R2 (early answerability):}
Before generating any code, the system must check that the query maps to available fields and supported derivations, and reject it otherwise.

\subsection{Reliable multi-step composition}
\label{sec:loops}

Even when a query is answerable, existing approaches cannot confirm that the generated program follows the steps intended by the user.
Chain-of-thought prompting~\cite{wei2022chain} asks the LLM to generate intermediate reasoning steps before answering.
AutoIOT~\cite{shen2025autoiot} applies such decomposition before writing code, and ReAct~\cite{yao2022react} alternates reasoning, actions, and observations so the model can revise its plan. These approaches improve how a model works through a task, but the reasoning trace is free text, so it cannot be checked mechanically against the user's intent.

Our bradycardia query shows the risk. A correct program must (i) detect R-peaks, (ii) compute intervals between them, (iii) convert intervals to beats per minute, and (iv) flag time windows below 60\,bpm.
The program can still run while being wrong: it may apply the 60\,bpm threshold to raw voltage, average heart rate over the whole day and hide short episodes, or convert sample intervals to seconds with the wrong sampling rate.

\vspace{0.5em}
\noindent \textbf{R3 (inspectable computation):}
The system must express each answer as a list of computational steps, with each step's operations, inputs, and dependencies stated, so an automatic checker can verify the plan before it produces a result.

%% file: tex/design.tex
\section{Flash-Fusion}
\label{sec:flash}

Flash-Fusion answers natural-language (NL) questions over sensor and telemetry tables using inspectable query plans.
A \emph{query plan} is an ordered list of predefined operations, each with its arguments and dependencies on earlier steps.

Flash-Fusion addresses the three requirements introduced in Section~\ref{sec:challenges}, mapping each to a design decision.
\begin{enumerate}[leftmargin=*]
  \item \textbf{R1 (bounded context):} Use schema-only prompting; the planner sees field summaries while predefined operators access the full dataset.
  \item \textbf{R2 (early answerability):} Before planning, check that the request maps to available fields and supported derivations, and reject it otherwise.
  \item \textbf{R3 (inspectable computation):} Select from a fixed vocabulary of typed operators, validate their arguments and plan dependencies, and reuse compatible operator sequences with query-specific field and value linking.
\end{enumerate}

Unlike AutoIOT and ReAct, which leave field binding, operator selection, and ordering to one code-generation step, Flash-Fusion separates scope checking, plan construction, and reuse into stages with inspectable outputs.

\begin{figure}[t]
  \centering
  \includegraphics[width=1.0\linewidth]{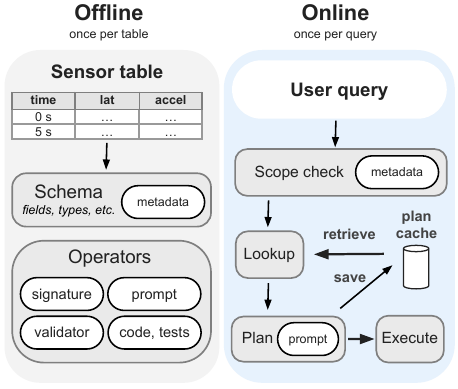}
  \caption{
    \textbf{Division of labor.}
    Offline, Flash-Fusion prepares a column profile from the schema and fixes the operator vocabulary.
    Online, it checks each query's scope, looks for a reusable plan skeleton, grounds a cached skeleton or constructs a new plan, and validates the plan before execution.
  }
  \label{fig:offline-online-split}
\end{figure}

\minihead{Offline preparation}
Flash-Fusion prepares two inputs before serving queries: a column profile describing the available table and an operator vocabulary defining the computations the planner may request (Figure~\ref{fig:offline-online-split}).

\minihead{Table and column profile}
Flash-Fusion receives sensor telemetry collected from edge devices and organizes each dataset (\eg a bus-analytics deployment) into a single typed, queryable table. During data preparation, incoming records are normalized into named fields and, where appropriate, aggregated into time windows, so every row has the same granularity: one reading or one window.
Each column represents a field (\eg \textit{timestamp}, \textit{latitude}, or \textit{accel\_variance}) with a logical data type: numeric, categorical, timestamp, boolean, or text.
This tabular representation provides a consistent interface for querying telemetry across different sensors and data sources.

The column profile is metadata computed from the table and is far smaller than the table, so it fits in the prompt.
For each column, it records the name, logical type, and number of distinct non-null values.
Numeric columns include their minimum, maximum, median, and standard deviation, computed over non-null values. Categorical columns include every distinct value with its count.

The current implementation loads one table into memory per query and does not ingest a continuous stream.
Flash-Fusion recomputes the column profile on every table load, so its fields, types, and values never go stale.
Only the column profile reaches the language model; operators execute over the full table.

\begin{figure}[t]
  \centering
  \includegraphics[width=\linewidth]{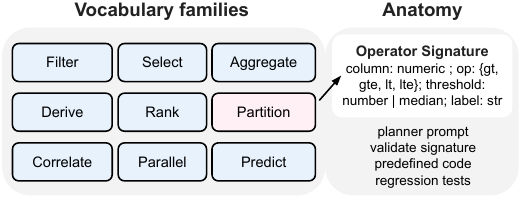}
  \caption{
    \textbf{Operator vocabulary and anatomy.}
    Left: 24 operators in 9 families.
    Right: each operator has a typed signature, planner-facing description, validator, predefined implementation, and regression tests.
    Query-time routing selects relevant families, reducing the planner prompt size.
  }
  \label{fig:vocab}
\end{figure}

\minihead{Operator vocabulary}
A column profile defines available fields but does not specify how the planner may use them.
Flash-Fusion therefore fixes a closed vocabulary of analytical operators before serving queries (Figure~\ref{fig:vocab}).
Together, these components separate the model's selection of an operator from the system's definition and execution of that operator.
For example, \textsc{SplitByThreshold} requires a column, comparator, threshold, and partition label.
Its threshold may be the symbolic value \texttt{median}, which the runtime resolves from the current table rather than requiring the model to retrieve the numeric value from the column profile.

The current vocabulary contains 24 operators in 9 families: filters (4), selections (1), aggregation (3), derived features (4), grouping and ranking (4), partition comparison (5), correlation (1), parallel branches (1), and a constrained predictive pipeline (1).
These families provide the units used to prune the planner's vocabulary at query time.

\begin{figure*}
  \centering
  \includegraphics[width=\textwidth]{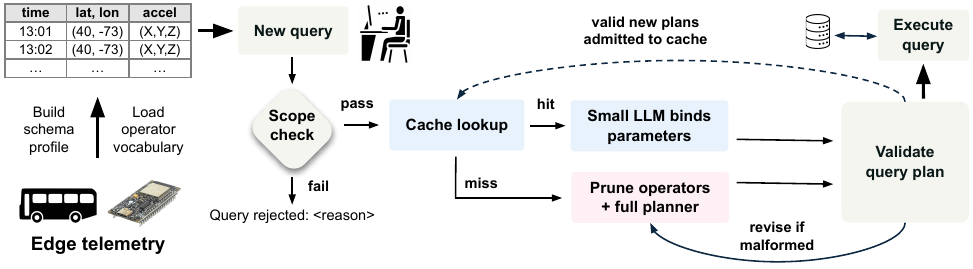}
  \caption{
    \textbf{Flash-Fusion Workflow.}
    A scope check rejects unsupported requests before plan construction.
    In-scope requests either retrieve and ground a compatible cached plan skeleton or construct a new plan from the pruned operator vocabulary.
    Both paths validate the resulting plan before executing it over the current table.
  }
  \label{fig:workflow}
\end{figure*}

\minihead{Online workflow and running example}
Flash-Fusion processes each query in four stages: scope check, cache lookup and grounding, planning on a cache miss, followed by execution.
Figure~\ref{fig:workflow} shows how the cache-hit and cache-miss paths converge on the same execution phase, running a validated query plan.

We follow one question through these stages:
\textit{``Is the northern half of the route rougher than the south?''}
The example uses a bus table containing 1{,}219 readings, including \textit{latitude} and \textit{accel\_variance}.
Answering the question requires interpreting ``northern half'' as a latitude-based partition and ``rougher'' through an acceleration-variance proxy, then comparing the two partitions.
The query does not name either field explicitly, so both mappings require semantic interpretation rather than literal column-name matching.

\subsection{Checking query scope}
\label{sec:firstcomponent}

\begin{figure}[t]
  \centering
  \includegraphics[width=\linewidth]{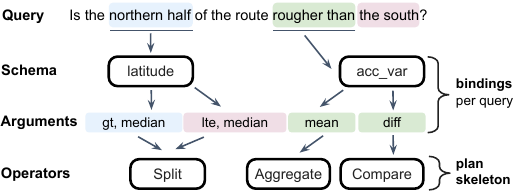}
  \caption{
    \textbf{From query concepts to executable bindings.}
    The query-to-schema rows link ``northern half'' to \textit{latitude} and ``rougher'' to the \textit{accel\_variance} proxy.
    The schema-to-argument rows make these choices explicit as a median latitude split, partition means, and a comparison.
    Scope checking establishes the query-to-schema mappings; planning or cache-hit grounding then supplies the arguments that parameterize each operator.
  }
  \label{fig:query-bindings}
\end{figure}

The first stage determines answerability (R2) before constructing an operator plan.
A request can be unsupported even when it resembles a familiar analytical task: the table may lack a required field, or the vocabulary may provide no supported derivation.
The scope checker therefore compares the query against the column profile and the available vocabulary, returning either an in-scope verdict or a rejection reason with concepts it could not resolve.

\minihead{Linking query concepts to fields}
The scope check begins the query-to-schema mapping shown in Figure~\ref{fig:query-bindings}.
For the running example, the column profile lists \textit{latitude} and \textit{accel\_variance} as numeric fields, so ``northern half'' can map to a latitude partition and ``rougher'' to an acceleration-variance comparison.
Each operator in the vocabulary (\eg Split, Aggregate, Compare) is a typed function over table columns: a mapped field supplies its input column, and the remaining arguments (\eg \textit{lte}, \textit{median}) fix its behavior.
At this stage, Flash-Fusion checks only that the required fields and derivations exist; later stages turn these mappings into operator arguments.

\minihead{Rejecting unsupported requests early}
Flash-Fusion rejects a request with a justification when a required concept has neither a supporting field nor a supported derivation from existing fields.
For example, a request for the bus stop with the most boardings is out-of-scope if the table contains neither boarding counts nor fields from which supported operators can derive them.
This rejection occurs before cache grounding or full planning, avoiding work on a request the system cannot answer.
The verdict guarantees grounding, not intent: an in-scope query binds only to existing fields and supported derivations, so plans cannot reference invented fields.
Among the grounded bindings, Flash-Fusion selects the one best supported by the column profile (\eg \textit{accel\_variance} for ``rougher'') and records any ambiguous concept alongside the plan, so the user can inspect the interpretation that was executed.
Verifying that a chosen binding matches user intent, for instance through interactive clarification, is orthogonal to answerability and left to future work.

\subsection{Looking up and grounding cached plans}
\label{sec:thirdcomponent}

The second stage avoids repeating operator selection and ordering when an earlier query already supplies a compatible plan.
Flash-Fusion maintains a \emph{plan cache} containing successfully executed queries and their \emph{plan skeletons}: ordered lists of operator identifiers, with duplicates preserved but arguments removed.
For the running example, the skeleton is \textsc{SplitByThreshold}, \textsc{SplitByThreshold}, \textsc{AggregatePartitions}, \textsc{ComparePartitions}.
The cache stores neither the answer nor previously bound columns, thresholds, or labels.

\minihead{Why reuse operator sequences}
Full planning requires the model to read operator descriptions, select operations, order their dependencies, and emit all arguments, which accounts for over 70\% of Flash-Fusion's end-to-end latency in our evaluation.
Queries that use the same operator sequence can avoid this composition work even when their wording or bound values differ.
For example, a paraphrase of the running query \textit{``Which side of the route is bumpier, north or south?''} has the same partition--aggregate--compare sequence.

Reusing a plan skeleton differs from directly retrieving the earlier answer produced by a similar query.
Semantic caches~\cite{bang2023gptcache} can return an earlier response, which need not remain valid for the current table.
Agentic Plan Caching (APC)~\cite{zhang2026agentic} instead extracts reusable templates from completed executions and adapts them to new requests.
Flash-Fusion follows this template-then-adapt approach but caches typed operator identifiers whose signatures are fixed offline.
It can therefore compare the reconstructed operator sequence to the cached plan skeleton and validate its new arguments against the current table.

\begin{figure}[t]
  \centering
  \includegraphics[width=\linewidth]{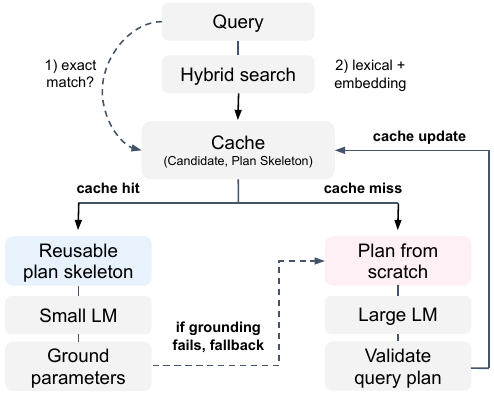}
  \caption{
    \textbf{Plan-cache lookup, grounding, and update.}
    An in-scope query first attempts an exact query-text match within its dataset, then lexical and embedding-based retrieval.
    An accepted candidate supplies a fixed plan skeleton for argument grounding.
    A miss, abstention, or unsuccessful reuse attempt routes to full planning; successful executions populate the cache.
  }
  \label{fig:cache_flow}
\end{figure}

\minihead{Finding a compatible skeleton}
Each entry retains the original query text, dataset identifier, plan skeleton, and metadata describing the requested computation.
Lookup first compares the new query text with cached query text within the same dataset, after trimming surrounding whitespace.
If no exact match exists, the matcher retrieves candidates using fuzzy lexical similarity and embedding cosine similarity based on an encoding-only transformer (Figure~\ref{fig:cache_flow}).
Lexical similarity favors shared wording, while embedding similarity also matches paraphrases with little wording overlap.

On the evaluated benchmarks, the matcher recovers 96.3\% of reusable queries with no false reuse (Section~\ref{sec:results}).
Matching is a heuristic, so we do not rely on it for correctness: a candidate is accepted only after its plan skeleton is compared against the reconstructed operator sequence and its arguments validate against the current table.
A candidate that fails either check, or a query with no sufficiently similar entry, falls back to full planning.

\minihead{Grounding a retrieved skeleton}
On a cache hit, the operator sequence is fixed, so a smaller language model fills arguments within the plan skeleton rather than selecting and ordering operators.
It receives the query, dataset name, table profile, and ordered operator list, and returns one parameter-binding list per step.
Flash-Fusion grounds these lists to the plan skeleton, reconstructing a typed query plan.
The model cannot add, remove, or reorder operators.

Figure~\ref{fig:query-bindings}'s schema and argument rows show this work for the running example.
The grounder binds \textit{latitude} to the two split steps, uses \texttt{gt} and \texttt{lte} with the \texttt{median} threshold, and assigns distinct partition labels.
It then binds \textit{accel\_variance} and \texttt{mean} to the aggregation and \texttt{diff} to the comparison.

The reconstructed plan then passes the same checks as a newly planned request (Section~\ref{sec:validation}).
Because this grounding task is constrained, a smaller model suffices; Table~\ref{tab:grounding_comparison} reports the resulting hit rate, latency, and cost.

\minihead{Populating the cache}
Flash-Fusion populates the cache online, without advance knowledge of future queries: when a cache miss is planned and executes successfully, it stores the query and its plan skeleton for later requests.

\subsection{Planning with the operator vocabulary}
\label{sec:cachemiss}

The third stage constructs a plan when no compatible plan skeleton is found in the cache.
After linking query concepts to fields (Section~\ref{sec:firstcomponent}), the planner must select operations and order them so each step consumes available inputs.
Flash-Fusion constrains this planning task to the operators defined offline, enabling inspectable computation (R3) and ensuring the model does not generate arbitrary code.

\minihead{Pruning the planner's vocabulary}
The planner prompt includes operator descriptions and worked examples, so supplying irrelevant operators increases input token overhead and the number of choices.
Flash-Fusion applies a deterministic router before full planning.
The router matches lexical cues, \eg ``compare'', ``above'', or ``correlation'', and returns the union of relevant operator families from Figure~\ref{fig:vocab}.

For the running query, the router retains 13 of 24 operators, including the three distinct operators required by the plan. It removes correlation, derived-feature, grouping and ranking, parallel-branch, and predictive families.
The operator vocabulary in the planner prompt shrinks by 79.5\%, which reduces one component of the planner's input, not its total token count; Section~\ref{sec:ablations} reports the end-to-end effect.

Pruning must preserve the operators needed to express a valid answer.
For queries that are long, ambiguous, contain unknown actions, or combine multiple intents, the router conservatively falls back to the full vocabulary rather than risk pruning an operator required for a valid plan.
This fall back trades prompt reduction for operator coverage when lexical cues prove insufficient.

\minihead{Constructing the running plan}
The full planner receives the query, table profile, selected operator definitions, worked examples, composition instructions, and structured response format.
It returns an ordered plan containing operator identifiers and their arguments.
For the running example, two \textsc{SplitByThreshold} steps define complementary latitude partitions, \textsc{AggregatePartitions} computes their mean acceleration variance, and \textsc{ComparePartitions} compares those means.
Table~\ref{tab:operator_trace} shows the resulting plan and execution trace.

\begin{table}[t]
\centering
\footnotesize
\setlength{\tabcolsep}{5pt}
\renewcommand{\arraystretch}{1.15}
\caption{\textbf{Example operator trace.}
Two splits define the latitude partitions; aggregation and comparison consume the state produced by earlier steps.
The runtime resolves the symbolic median value from the full table.
}
\label{tab:operator_trace}
\begin{tabularx}{\columnwidth}{@{}l X@{}}
\toprule
\textbf{Operator} & \textbf{Grounded execution} \\
\midrule
\textsc{SplitByThreshold}
  & \textit{latitude} $>$ median $=33.7767$, north (609 rows) \\

\textsc{SplitByThreshold}
  & \textit{latitude} $\leq$ median $=33.7767$, south (610 rows) \\

\textsc{AggregatePartitions}
  & mean(\textit{accel\_variance}) for each partition \\

\textsc{ComparePartitions}
  & difference
    $\rightarrow$ \textbf{northern higher} \\
\bottomrule
\end{tabularx}
\end{table}

\subsection{Validating and executing plans}
\label{sec:validation}

The final stage checks that either path produced a query plan suitable for the current table.
Using its plan validator, Flash-Fusion validates both the plan's operator structure and arguments before executing any analytical operator.

\minihead{Checking bindings against the current table}
The binding checks confirm that referenced columns exist, their types suit the operators that use them, and named categorical values appear in the data.
For the running example, \textit{latitude} and \textit{accel\_variance} must be numeric for the median split and mean aggregation.
These checks apply to both newly planned and cache-grounded query plans.
They complement the scope check (Section~\ref{sec:firstcomponent}), which evaluates whether the query has a supported interpretation, while this stage checks the concrete arguments produced for that interpretation.

\minihead{Checking operator structure and dependencies}
The plan validator checks the language model's output against each operator's schema.
The plan must be a nonempty, ordered list whose steps name known operators, contain no extra fields, and satisfy parameter-type and cross-field constraints.
For example, a numeric binning step requires a positive width, and temporal binning over numeric timestamps requires an epoch unit.

Composition checks then verifies that each step consumes only state produced by earlier steps.
In the running example, the northern and southern partitions must exist before aggregation, and both means must exist before comparison.
Placing \textsc{ComparePartitions} before \textsc{AggregatePartitions} fails validation because the required aggregate values are unavailable.
For a cache-grounded plan, the system additionally checks that its ordered operator identifiers exactly match the retrieved skeleton.

\minihead{Executing over the full table}
After validation, Flash-Fusion invokes the selected operators and their arguments over the current table.
The runtime computes data-dependent arguments, such as the median, directly from the table.
For the running example (Table~\ref{tab:operator_trace}), the median-latitude split produces 609 northern and 610 southern readings.
Their mean acceleration variances are 0.3703 and 0.1631, respectively, giving a difference of 0.2072.
Flash-Fusion answers that the northern half is rougher and reports the proxy (\textit{accel\_variance}), the median-latitude split, and both means.

%% file: tex/implementation.tex
\section{Flash-Fusion Implementation}
\label{sec:ff-implementation}

Flash-Fusion is implemented as a Python package of about 30,000 lines of code.
We build on existing libraries and state the role of each below. All experiments in Section~\ref{sec:results} run on a single local machine, with language model inference served remotely.

\minihead{Offline data preparation}
Flash-Fusion must turn uploaded telemetry into a typed table and give the planner only its summary.
A user submits a CSV or text file, which we load into an in-memory Pandas~\cite{mckinney2011pandas} DataFrame.
Flash-Fusion then normalizes records into named, typed columns, aggregates them into time windows where appropriate, and computes the column profile described in Section~\ref{sec:flash}.

The query plan must also be represented in a form the system can execute.
We define each operator as a Pydantic~\cite{colvin2024pydantic} model, a Python class that declares its fields, their types, and cross-field rules, and tag it with a fixed operator name. Pydantic uses the tag to parse each plan step in the language model's JSON output against that operator's schema alone, so an unknown operator, an undeclared field, or a violated rule fails at parse time, before any code runs.

\minihead{Online planning and caching}
Language model calls fall into two paths. On a cache miss, Qwen3-Max~\cite{qwen3maxblog} plans the full query. On a cache hit, Ministral-8B~\cite{ministralblog} grounds the retrieved plan skeleton in the new query; Table~\ref{tab:grounding_comparison} lists the other small grounding models we evaluated. Every baseline uses Qwen3-Max for planning and code generation, so performance differences reflect system design and not model capability. All calls, for Flash-Fusion, its ablations, and the baselines, go through OpenRouter~\cite{openrouter_pricing}, which reports the token counts and billed cost we use.
LangChain~\cite{langchain_errors_2026} handles message formatting and call logging, and its DataFrame agent provides the ReAct~\cite{yao2022react} baseline.

The plan cache is a JSON registry whose entries hold the query text, dataset identifier, and plan skeleton. Lookup first checks for an exact text match within the dataset. If none exists, it retrieves candidates in two ways: (a) lexical matching, using RapidFuzz~\cite{ye2021rapidfuzz} for fuzzy string similarity after Snowball stemmer~\cite{snowball} normalizes terms; and (b) dense matching, using cosine similarity between embeddings from the all-MiniLM-L6-v2 encoder~\cite{yin2024study}, loaded from HuggingFace~\cite{huggingface} via sentence-transformers. We chose these components because they run in-process on a CPU.

\minihead{Experimental setup}
Latency depends on where each stage runs, so we separate local from remote work.
Local stages run on a MacBook Air (Apple M4, 16\,GB RAM, macOS 26.5.1, Python 3.11.13). These stages are data loading, schema profiling, cache lookup (including the embedding model on CPU), operator validation, and plan execution.
Remote stages are all language model inference, which OpenRouter serves over home Wi-Fi ($\sim$90\,Mbps each way, 25\,ms mean round-trip time, no packet loss). We measure end-to-end latency with wall-clock timers around each query, so it includes network round trips and provider-side queuing.
We reproduce each baseline (Section~\ref{sec:baselines}) from its released code, and we test operators, validation, routing, and the cache with Pytest.

%% file: tex/evaluation.tex
\section{Evaluation}
\label{sec:evaluation}

Our evaluation aims to answer the following research questions:
\begin{enumerate}[leftmargin=*, itemsep=1pt]
  \item Is Flash-Fusion more accurate than existing systems across direct, reasoning, predictive, and out-of-scope queries?
  \item Does it keep latency interactive across query types?
  \item Is it cheaper than the baselines?
  \item How much does each component (caching, pruning, and planner instructions) contribute (Section~\ref{sec:ablations})?
\end{enumerate}

\subsection{Datasets}
\label{sec:datasets}

We evaluate all systems on three datasets of varying scale and modality; WISDM and MIT-ECG were also used to evaluate AutoIOT~\cite{shen2025autoiot}.

\minihead{WISDM}
A standard human activity recognition dataset~\cite{weiss2019wisdm} comprising approximately 1.1 million rows of tri-axial accelerometer data from 51 subjects.

\minihead{MIT-ECG}
The MIT-BIH Arrhythmia dataset~\cite{apandi2018arrhythmia} provides 26 million rows of high-frequency electrocardiogram data, stressing scalability.

\minihead{Bus Telematics}
We collected telemetry on a university bus fleet's highest-traffic route over 2 hours and 14 minutes, yielding 2,648 rows of GPS
coordinates and windowed tri-axial accelerometer statistics.
It is the smallest dataset by design, exercising GPS-grounded spatial queries; we will release it upon acceptance.

\subsection{Query taxonomy \& scoring}
\label{sec:taxonomy}

We evaluate Flash-Fusion using query types selected to cover evaluation dimensions from established tabular QA and text-to-SQL benchmarks.
TableBench~\cite{wu2025tablebench} evaluates table reasoning through fact-checking, numerical reasoning, data analysis, and visualization, while BIRD~\cite{li2023can} differentiates question--SQL instances by difficulty to expose failures in schema grounding and multi-step reasoning. We organize the benchmark into four query categories aligned with the challenges in Section~\ref{sec:challenges}: \emph{direct}, \emph{reasoning}, \emph{predictive}, and \emph{out-of-scope}.

\begin{itemize}[leftmargin=*, nosep]
  \item \textbf{Direct (Q1--Q4):} Simple filtering or aggregation (\eg \textit{``How many total samples are classified as Walking?''}).
  \item \textbf{Reasoning (Q5--Q12):} Derived-feature proxies and multi-step decomposition (\eg \textit{``Identify the user whose duration of stationary activities exceeds their duration of active locomotion by the largest margin.''}).
  \item \textbf{Predictive (Q13--Q16):} Trend forecasting (\eg \textit{``Predict whether an annotation is present for the first row in the holdout set.''}), mirroring AutoIOT's~\cite{shen2025autoiot} benchmark.
  \item \textbf{Out-of-scope (Q17--Q20):} Requests unsupported by the schema (\eg \textit{``Are female subjects more likely to have a higher cadence during stair climbing?''}), each categorized by violation type: missing attribute, unsupported inference, or out-of-domain forecasting.
\end{itemize}

We design 20 queries per dataset (60 total), each assigned an operation tag (\eg \textsc{Filter+Aggregate}). Each query is paraphrased twice, yielding 180 queries that test whether the plan cache recognizes rephrased queries.

Ground-truth answers for direct, reasoning, and predictive queries are computed deterministically against the full dataset.
We score each answer with an LLM judge, a paradigm validated in MT-Bench~\cite{zheng2023judging}: 1 if it matches the ground truth and 0 otherwise, with no partial credit.
For out-of-scope queries, the binary policy is inverted: any response that does not explicitly decline to answer scores 0, consistent with the abstention protocol in SQuAD 2.0~\cite{rajpurkar2018know}.
We run each benchmark three times in randomized order and report the mean with min--max error bars.

\minihead{Metrics}
\begin{itemize}[leftmargin=*]
  \item \textit{Accuracy:} percentage of evaluation queries for which the system returns a correct answer. Higher is better.
  \item \textit{Latency:} wall-clock time from query submission to complete response, including all pipeline stages, agent thought-action loops, and execution overhead. Lower is better.
  \item \textit{Cost:} cumulative input/output token cost across all API calls required to service a single query, computed from the API provider's~\cite{openrouter_pricing} pricing. Lower is better.
\end{itemize}

\subsection{Baselines}
\label{sec:baselines}

We compare against four baselines that represent the main paradigms for applying LLMs to sensor data.

\textbf{ReAct}~\cite{yao2022react} converts a natural-language query directly into Python code executed over the dataset. It is the simplest baseline and, with today's LLM coding ability, the strongest, but cannot detect out-of-scope queries; our only change is a prompt telling it to decline unanswerable queries.

\textbf{AutoIOT}~\cite{shen2025autoiot} generates code via background-knowledge retrieval, program synthesis, and an iterative self-correction loop (up to 5 rounds) using compiler error feedback. We implement it as originally presented.

\textbf{HARGPT}~\cite{ji2024hargpt} formats IoT readings directly as text and queries the model with zero-shot chain-of-thought reasoning. We pre-filter by the query's targeted entity and subsample to a 1,000--2,500 row window, fitting the data into a single 15,000--25,000 token prompt to respect context window and API burst limits~\cite{groqpricing}.

\textbf{LLMSense}~\cite{ouyang2024llmsense} breaks long sensor streams into temporal windows; an edge LLM narrates each chunk (capped at 160 rows) and a cloud LLM aggregates the narrations to reason over the full dataset.

\subsection{Results}
\label{sec:results}

The following parts directly address the first three research questions listed at the start of Section~\ref{sec:evaluation}.

\begin{figure}[t]
  \centering
  \includegraphics[width=1.0\linewidth]{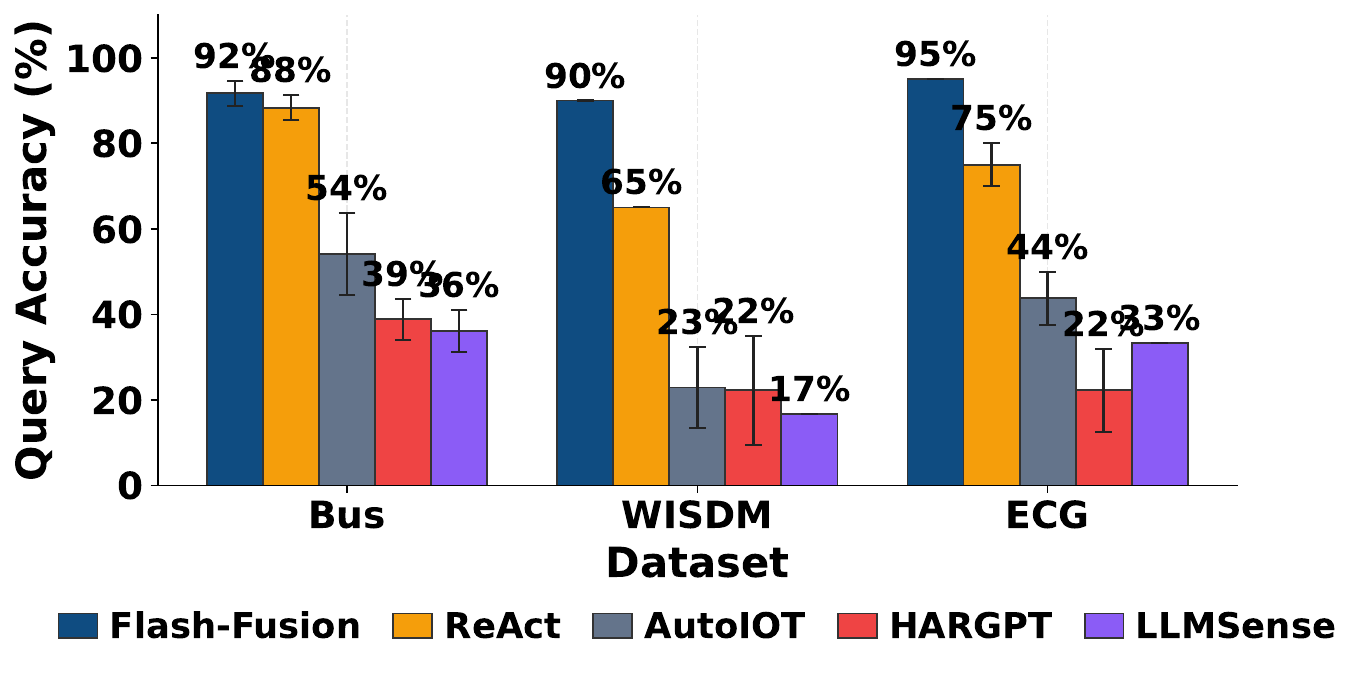}
  \caption{\textbf{Query Accuracy Across Baselines.}
  Flash-Fusion is the most accurate system on every dataset.}
  \label{fig:query_accuracy}
\end{figure}

\minihead{Flash-Fusion is more accurate and consistent}
Figure~\ref{fig:query_accuracy} compares mean accuracy per dataset. Averaged over datasets, Flash-Fusion achieves 92.2\% accuracy, 16.1 points above the strongest baseline, ReAct (76.1\%), and is over 4$\times$ more consistent run-to-run.

\begin{figure}[t]
  \centering
  \includegraphics[width=1.0\linewidth]{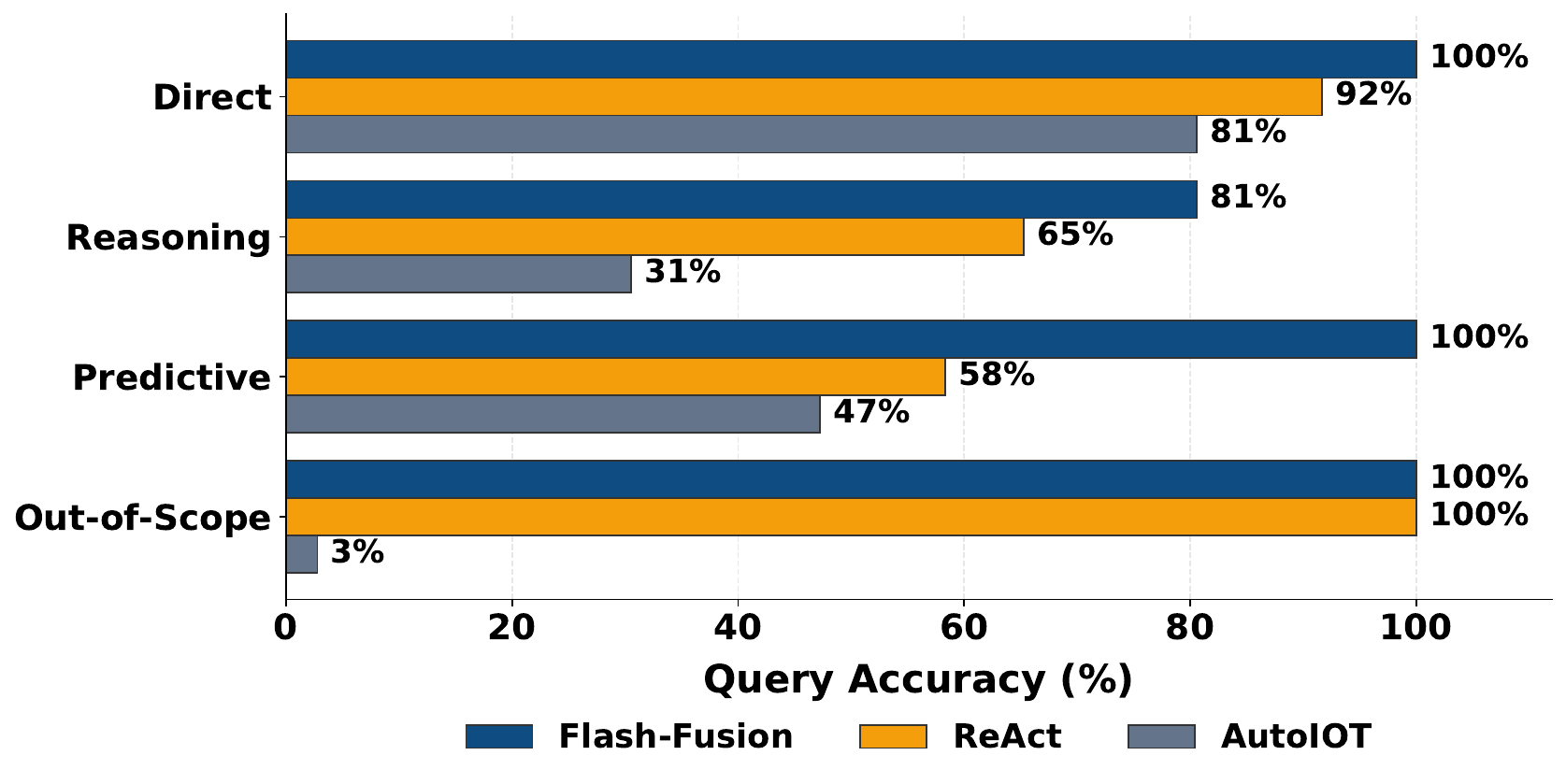}
  \caption{\textbf{Query Accuracy Across Types.}
  Flash-Fusion is accurate across categories, while ReAct struggles on reasoning and predictive queries. AutoIOT lacks an abstention protocol and fails to reject out-of-scope queries.}
  \label{fig:query_accuracy_by_type}
\end{figure}

The query type results show where this accuracy gap arises (Figure~\ref{fig:query_accuracy_by_type}). ReAct's code-generation and abstention approach achieves >95\% accuracy on direct and out-of-scope queries, but <70\% on reasoning and predictive queries.
These harder queries require multi-hop probes, for which ReAct's zero-shot approach must rediscover derived-feature proxies on every run.
Re-deriving the execution path across runs can produce different, sometimes incorrect paths, explaining its high run-to-run variance.
In contrast, Flash-Fusion answers queries with a closed set of operator families (Figure~\ref{fig:vocab}), constraining its planner's search space. This yields higher, consistent accuracy across queries involving multiple steps.
Flash-Fusion fails on these queries when the cache contains no reusable plan skeleton and its fall back planning stage cannot construct one from scratch.

AutoIOT~\cite{shen2025autoiot} likewise achieves >80\% accuracy on direct queries but <50\% on reasoning and predictive queries, despite its retrieval, drafting, and compiler-feedback refinement loop. It also struggles with queries requiring derived-feature proxies and with identifying out-of-scope queries.

HARGPT~\cite{ji2024hargpt} and LLMSense~\cite{ouyang2024llmsense} also struggle with queries requiring exact values from the underlying data, such as \textit{``What is the maximum recorded x-acceleration for user 15?''}.

Higher accuracy must be achieved within low latency and spend, which we evaluate next.

\minihead{Constrained planning cuts latency}
We first compare end-to-end latency across query types, then break it down by stage to identify where the difference comes from.

\begin{figure}[t]
\centering
\includegraphics[width=1.0\linewidth]{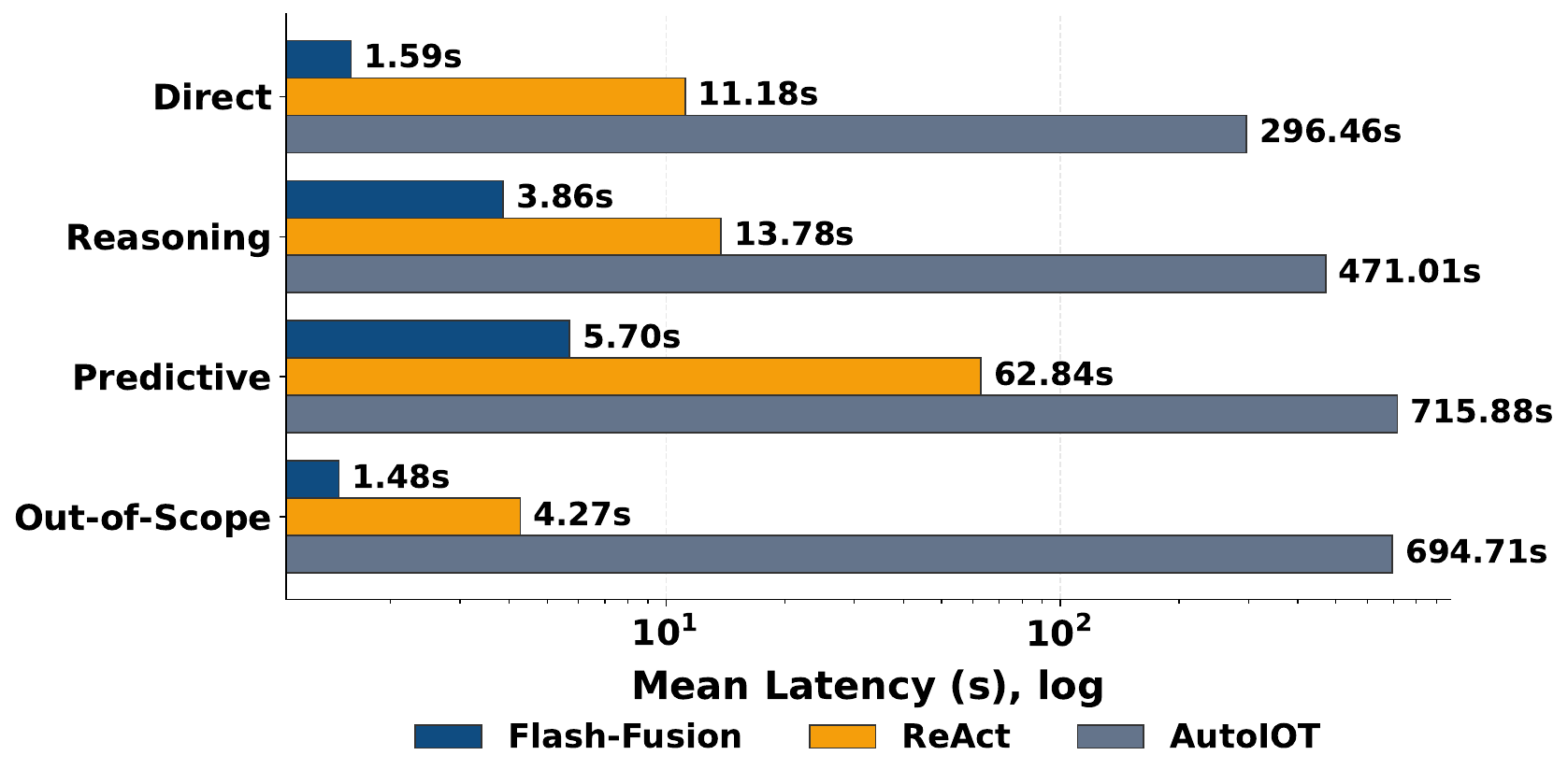}
\caption{\textbf{Mean Latency Across Query Types.}
Flash-Fusion has the lowest mean latency for every query type (log scale).}
\label{fig:latency_types}
\end{figure}

Figure~\ref{fig:latency_types} reports mean latency per query type for Flash-Fusion and the two most accurate baselines.
Flash-Fusion has the lowest latency in every category because of two design choices: the planner composes programs from a constrained set of typed operators, and it reuses plan skeletons on a cache hit.
Both choices shrink the planner's search space and the number of tokens it generates.
ReAct, in contrast, generates code from an open-ended action space with no reusable structure.
AutoIOT's repeated self-correction loops raise its latency further.

\begin{figure}[t]
\centering
\includegraphics[width=1.0\linewidth]{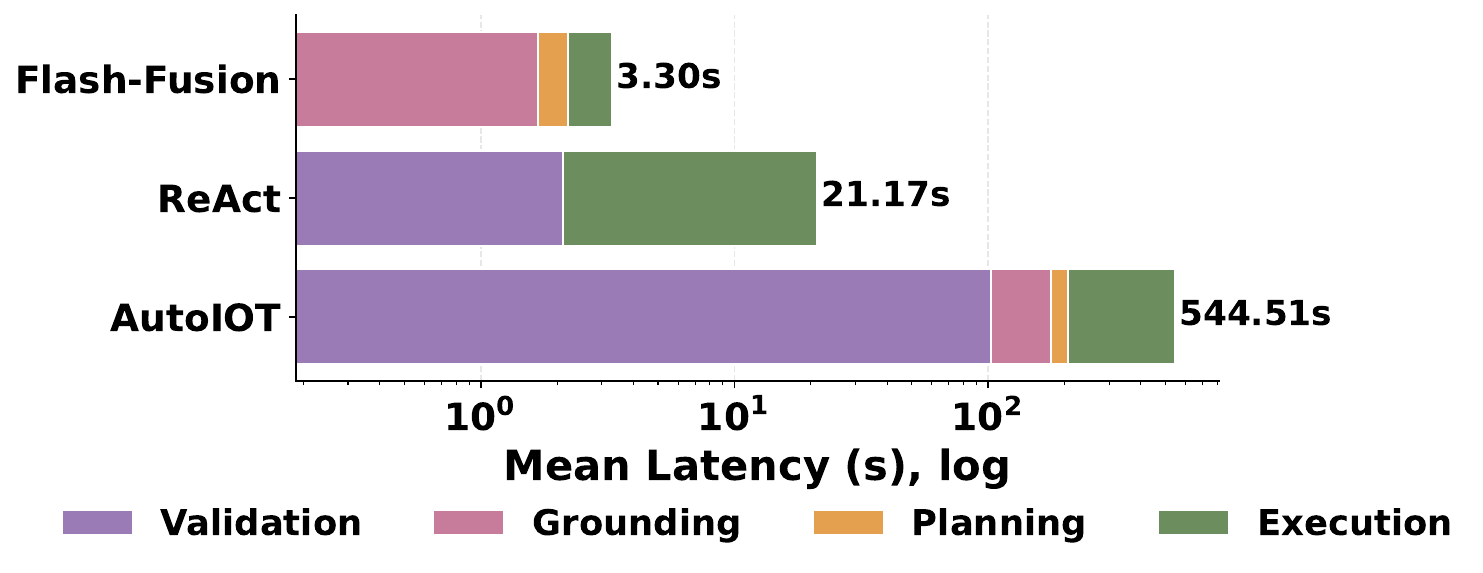}
\caption{\textbf{Latency by Stages Across Baselines.}
On similar queries, Flash-Fusion grounds values into a cached plan skeleton with a smaller model, so its stages stay short.}
\label{fig:latency_by_stages}
\end{figure}

To identify the key bottlenecks, we break latency into four stages (Figure~\ref{fig:latency_by_stages}): validation, the scope check of Section~\ref{sec:firstcomponent}; grounding of a cached skeleton (Section~\ref{sec:thirdcomponent}); planning (Section~\ref{sec:cachemiss}); and execution (Section~\ref{sec:validation}).

Flash-Fusion spends less time than the baselines on each stage it runs. On a cache hit, Flash-Fusion skips planning and only grounds values into the retrieved plan skeleton.
A smaller model does this cheaply because its context is scoped and its output is short.
Validation is barely visible in Figure~\ref{fig:latency_by_stages} because of the log-scale x-axis, but the stage-wise ablation (Figure~\ref{fig:ablation_latency}) shows it clearly.
Planning runs only on a cache miss, when the smaller model grounds values incorrectly, validation fails, and Flash-Fusion falls back to full planning.

The baselines spend their time in different places.
ReAct has no separate planning stage: after its scope check, it interprets the query and generates code in one open-ended execution step, which dominates its latency because nothing constrains or reuses its reasoning.
AutoIOT spends most of its time in validation, repeatedly retrieving knowledge and refining the query.

\begin{table}[t]
\centering
\footnotesize
\setlength{\tabcolsep}{4pt}
\renewcommand{\arraystretch}{1.15}
\caption{\textbf{Comparison Against Baselines on a Per-Query Basis.}
Flash-Fusion achieves the highest mean accuracy with the lowest latency and cost.}
\label{tab:system_comparison}

\begin{tabular}{lcccc}
\toprule
\textbf{System}
& \textbf{Accuracy (\%)}
& \textbf{Latency (s)}
& \textbf{Cost (\$ $\times 10^{-5}$)} \\

\midrule

\rowcolor{flashfusion!20}
\textbf{Flash-Fusion}
& \textbf{92.22 $\pm$ 0.96}
& \textbf{3.30 $\pm$ 0.28}
& \textbf{24.21 $\pm$ 4.00} \\

ReAct~\cite{yao2022react}
& 76.11 $\pm$ 2.55
& 21.17 $\pm$ 0.22
& 114.41 $\pm$ 19.95 \\

AutoIOT~\cite{shen2025autoiot}
& 40.28 $\pm$ 5.24
& 544.51 $\pm$ 65.32
& 1{,}708.50 $\pm$ 90.92 \\

HARGPT~\cite{ji2024hargpt}
& 27.78 $\pm$ 2.78
& 165.08 $\pm$ 65.28
& 5{,}567.29 $\pm$ 172.02 \\

LLMSense~\cite{ouyang2024llmsense}
& 25.69 $\pm$ 1.20
& 75.12 $\pm$ 0.78
& 8{,}330.76 $\pm$ 125.09 \\

\bottomrule
\end{tabular}
\end{table}

\minihead{Query cost across systems}
In addition to being accurate and fast, Flash-Fusion is also cheap (Table~\ref{tab:system_comparison}), followed by ReAct, the least expensive baseline. ReAct processes the query and generates a program that answers it, whereas the other baselines introduce additional token-intensive operations: AutoIOT's grounding stage comprises several rounds of background knowledge retrieval and query refinement, HARGPT includes the dataset directly in its context, and LLMSense summarizes chunks of sensor streams. Flash-Fusion instead reuses cached plan skeletons and grounds their parameters with a smaller model, reducing its reliance on full planning for repeated query patterns.

\minihead{Partial cache reuse is sufficient for cost savings}
The aggregate comparison does not show how much cache reuse Flash-Fusion needs to become cheaper than ReAct.
To isolate this effect, we simulate expected query cost over cache-hit rates from 0\% to 100\%, averaging across 50 seeded, query-type-balanced assignments. Cache hits use a smaller grounding model, while cache misses fall back to Qwen3-Max for full planning (Section~\ref{sec:ff-implementation}).

\begin{figure}[t]
\centering
\includegraphics[width=1.0\linewidth]{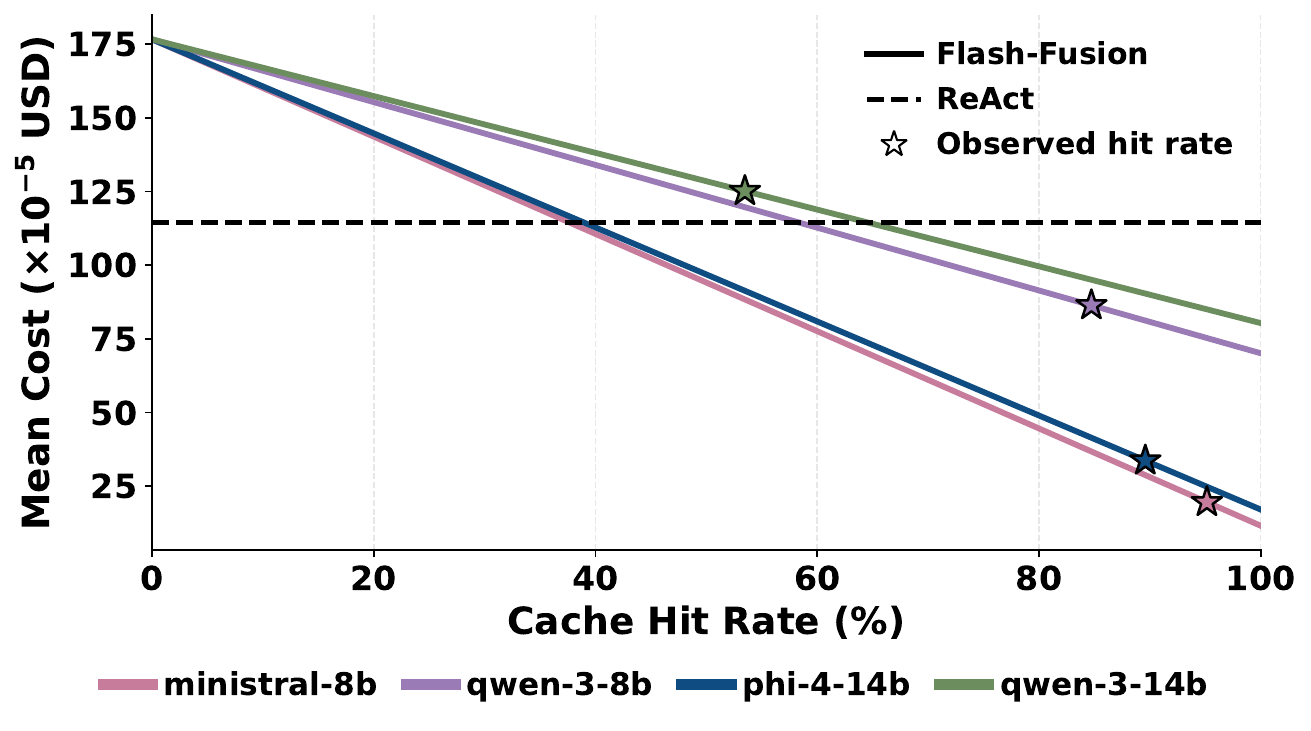}
\caption{\textbf{Cache-Hit Rate Versus Query Cost.}
Flash-Fusion delivers cost savings across several grounding models, falling below ReAct's average query cost at modest cache hit rates.
The dashed line marks ReAct (Qwen3-Max).
Stars mark measured hit rates from Table~\ref{tab:grounding_comparison}; segments between the shared endpoint and each star are linear interpolations, while segments beyond the stars are extrapolated estimates.}
\label{fig:cache_cost}
\end{figure}

Figure~\ref{fig:cache_cost} shows the resulting cost projections, with ReAct's mean query cost as a horizontal reference. All Flash-Fusion lines share the same 0\% endpoint, where every query requires full planning.
Higher hit rates route more queries to the cheaper grounding path, lowering mean cost.

\minihead{Grounding model determines achievable reuse}
The sensitivity analysis (Figure~\ref{fig:cache_cost}) establishes the hit rates needed for savings, but grounding models differ in their ability to achieve those rates. Table~\ref{tab:grounding_comparison} compares their measured cache-hit rates, latency, and cost. We select Ministral-8B because it achieves the highest measured hit rate (95.14\%) and the lowest mean latency and cost among the evaluated models.

\begin{table}[t]
\centering
\footnotesize
\setlength{\tabcolsep}{4pt}
\renewcommand{\arraystretch}{1.15}
\caption{\textbf{Comparison Against Grounding Models during Cache Hits.}
Ministral-8B gives the highest hit rate and the lowest latency and cost.
}
\label{tab:grounding_comparison}

\begin{tabular}{lcccc}

\toprule

\textbf{Model $\uparrow$}
& \textbf{Cache Hit Rate (\%)}
& \textbf{Latency (s)}
& \textbf{Cost (\$ $\times 10^{-5}$)} \\

\midrule

llama-3.2-1b
& 13.19 $\pm$ 8.42
& 29.87 $\pm$ 26.25
& 117.96 $\pm$ 47.03 \\

llama-3.2-3b
& 18.06 $\pm$ 7.89
& 6.59 $\pm$ 0.32
& 152.42 $\pm$ 18.61 \\

gemma-3-4b
& 29.17 $\pm$ 4.17
& 9.88 $\pm$ 0.91
& 74.97 $\pm$ 20.10 \\

\rowcolor{flashfusion!20}
\textbf{ministral-8b}
& \textbf{95.14 $\pm$ 1.20}
& \textbf{3.82 $\pm$ 0.58}
& \textbf{19.62 $\pm$ 1.76} \\

qwen-3-8b
& 84.72 $\pm$ 4.81
& 19.05 $\pm$ 3.14
& 86.37 $\pm$ 15.38 \\

llama-3.1-8b
& 50.69 $\pm$ 4.81
& 9.49 $\pm$ 3.10
& 148.26 $\pm$ 34.82 \\

granite-4.2-8b
& 14.58 $\pm$ 7.22
& 7.74 $\pm$ 0.48
& 177.69 $\pm$ 21.43 \\

gemma-3-12b
& 67.36 $\pm$ 12.21
& 5.65 $\pm$ 0.77
& 70.40 $\pm$ 18.74 \\

phi-4-14b
& 89.58 $\pm$ 3.61
& 4.89 $\pm$ 0.72
& 33.71 $\pm$ 8.89 \\

qwen-3-14b
& 53.47 $\pm$ 15.36
& 13.94 $\pm$ 2.59
& 125.11 $\pm$ 45.85 \\

\bottomrule

\end{tabular}

\end{table}

\subsection{Ablations}
\label{sec:ablations}

We remove components cumulatively: \textit{- cache} removes cache-based grounding; \textit{- prune} additionally removes pruning; and \textit{- prompt} additionally removes the planner's instructions and worked examples.

\begin{figure}[t]
\centering
\includegraphics[width=1.0\linewidth]{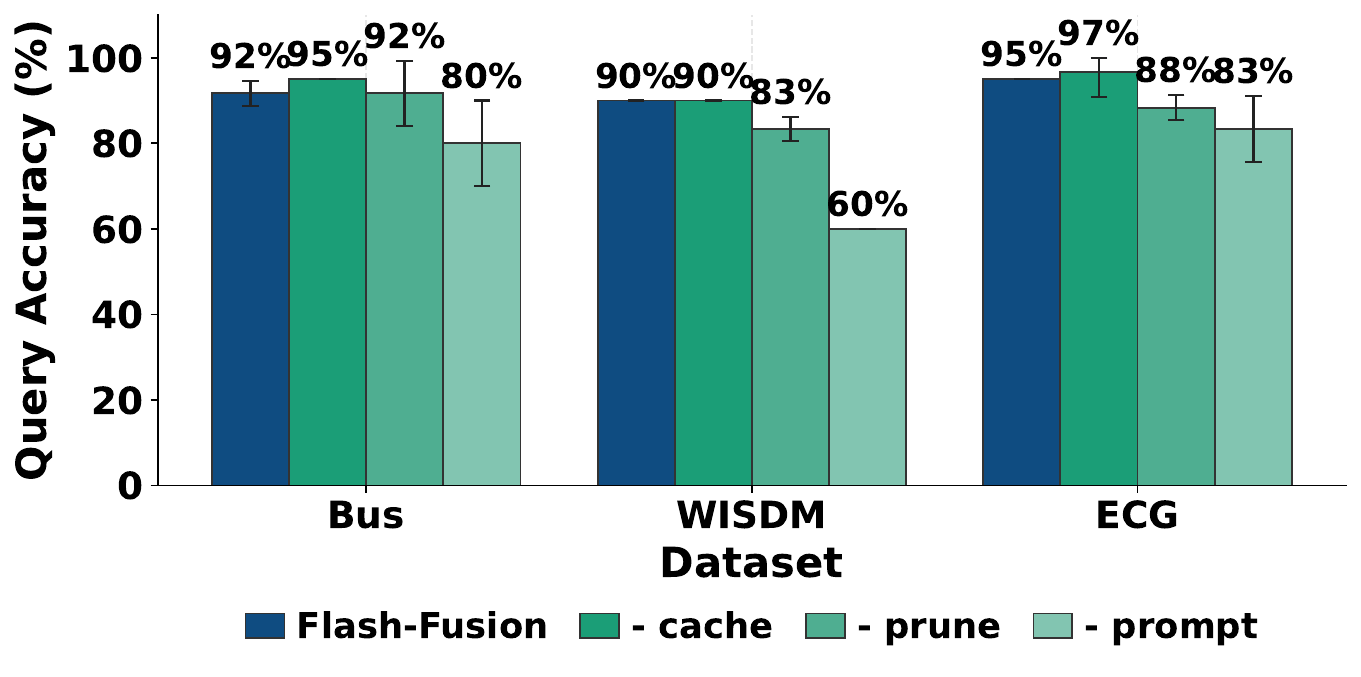}
\caption{\textbf{Query Accuracy Across Ablations.} Pruning and prompt instructions improve accuracy; removing the cache barely changes it because the full planner handles every query.}
\label{fig:ablation_accuracy}
\end{figure}

\minihead{Each component contributes differently}
Flash-Fusion's accuracy is the most consistent across datasets (Figure~\ref{fig:ablation_accuracy}). Without the cache, accuracy barely changes (93.9\% vs.\ 92.2\%) because the full planner then handles every query.
In contrast, ablating operator pruning reduces accuracy. Pruning restricts the planner's operator families to a query-relevant subset, improving answer accuracy, rather than merely reducing the planner's search space.
Removing prompt instructions and worked examples produces the largest accuracy drop, highlighting the importance of in-context learning while constructing plans within domain-specific language (DSL) constraints.

\begin{figure}[t]
\centering
\includegraphics[width=1.0\linewidth]{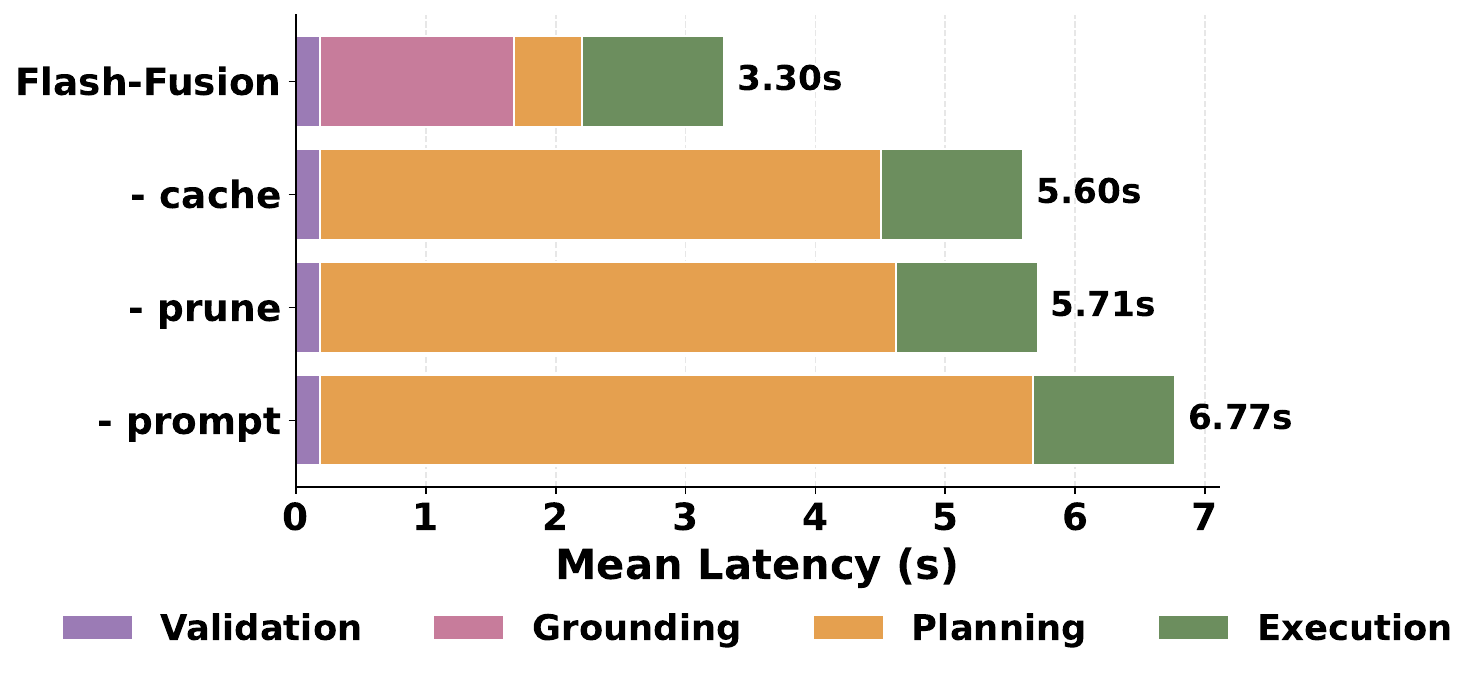}
\caption{\textbf{Latency by Semantic Stage Across Ablations.}
Planning dominates latency without the cache, pruning, or instruction prompts.}
\label{fig:ablation_latency}
\end{figure}

Flash-Fusion has the lowest latency among its ablations (Figure~\ref{fig:ablation_latency}), showing that each component contributes to the speedup: the cache avoids planning except on a miss, pruning leaves fewer operators to consider, and removing instructions makes planning slowest.

The cache has the largest effect on cost (Figure~\ref{fig:cache_cost}; Table~\ref{tab:ablation_comparison}). \textit{- prune} costs slightly more than \textit{- cache} because more input tokens reach the planner, while \textit{- prompt} costs slightly less because its prompt is shorter.

\begin{table}[t]
\centering
\footnotesize
\setlength{\tabcolsep}{4pt}
\renewcommand{\arraystretch}{1.15}
\caption{\textbf{Comparison Against Ablations.}
Flash-Fusion has the most consistent accuracy and the lowest latency and cost.}
\label{tab:ablation_comparison}

\begin{tabular}{lcccc}
\toprule
\textbf{Ablation}
& \textbf{Accuracy (\%)}
& \textbf{Latency (s)}
& \textbf{Cost (\$ $\times 10^{-5}$)} \\

\midrule

\rowcolor{flashfusion!20}
\textbf{Flash-Fusion}
& \textbf{92.22 $\pm$ 0.96}
& \textbf{3.30 $\pm$ 0.28}
& \textbf{24.21 $\pm$ 4.00} \\

$\setminus$ cache
& 93.89 $\pm$ 1.92
& 5.60 $\pm$ 0.10
& 186.65 $\pm$ 8.90 \\

$\setminus$ prune
& 87.78 $\pm$ 4.19
& 5.71 $\pm$ 0.27
& 190.20 $\pm$ 0.71 \\

$\setminus$ prompt
& 74.44 $\pm$ 4.81
& 6.77 $\pm$ 0.14
& 167.48 $\pm$ 7.63 \\

\bottomrule
\end{tabular}
\end{table}

%% file: tex/relatedwork.tex
\section{Related Work}

Flash-Fusion integrates three research domains: (1) \textbf{semantic query processing}, which exposes domain-specific language operators over data; (2) \textbf{language models for sensor analysis}, which connect sensor signals to language; and (3) \textbf{agent workflow memory}, which reuses prior plans to reduce the cost of execution.

\subsection{Semantic query processing}

Existing semantic query engines let users express analytical operations in natural language, but they pay for an LLM call on most records.
LOTUS defines semantic filters, joins, maps, rankings, and aggregations, and provides statistical accuracy guarantees relative to a reference algorithm~\cite{patel2025semantic}.
Palimpzest and Abacus choose among models and physical implementations to trade off cost, latency, and quality~\cite{liu2025palimpzest, russo2025abacus}.
SemBench evaluates such engines on tables combined with text, image, and audio data~\cite{lao2025sembench}. Because these engines interpret each record or record pair with an LLM, optimization focuses on reducing and approximating those calls.

Flash-Fusion shares the separation between a declarative request and its execution, but it first parses a user's natural-language request into an executable query plan. The LLM produces a plan over a closed vocabulary of typed operators, grounded in the dataset's schema.

\subsection{Language models for sensor analysis}

Language models can learn representations that enable them to interpret sensor signals directly. SensorLM aligns wearable data with captions at the scale of 59.7 million hours~\cite{zhang2026sensorlm}. IoT-LM adapts pretrained LLMs to 12 sensor modalities~\cite{mo2024iot}. LLaSA couples an IMU encoder with an LLM for open-ended activity question answering~\cite{asif2025llasa}. HoloLLM targets less common modalities such as LiDAR and mmWave radar~\cite{zhou2026holollm}.

Flash-Fusion is complementary: it assumes the measurements already sit in a table and learns no sensor encoder. The LLM selects the columns and operations that answer the query, and the underlying execution engine computes the numbers. A learned model could supply derived attributes that Flash-Fusion's operators then consume.

\subsection{Agent workflow memory}

Agent memory systems reuse prior experiences, but they operate on free-form trajectories, so a reused plan still needs adaptation and may fail. Agent Workflow Memory induces reusable workflows from earlier trajectories and supplies them to later generations~\cite{wang2024agent}. Agentic Plan Caching stores plan templates from completed executions, matches new requests by keyword, and adapts each template with a small model~\cite{zhang2026agentic}. AgentStop targets efficiency differently: it predicts from execution signals which runs will fail and terminates them early~\cite{pham2026agentstop}.

Flash-Fusion's skeleton cache reuses the operator sequence of an eligible earlier plan. The cache re-grounds the plan's parameters against the current schema and applies the same validation gates before executing on current data. Flash-Fusion rejects unsupported requests before execution instead of stopping them partway through. The open-ended ReAct agent remains as a fall back for queries the operator vocabulary cannot express.

%% file: tex/conclusion.tex
\section{Conclusion}
\label{sec:conclusion}

Natural-language analytics over IoT sensor data is unreliable with existing LLM approaches.
Raw-prompting exhausts the context window and loses the numeric detail that analytical queries require, while unconstrained code-synthesis agents hallucinate answers when the schema cannot support a query and let one faulty intermediate step derail a multi-step query.

We presented Flash-Fusion, an agentic framework that addresses both failure modes: it checks that available fields and supported derivations can answer a query, builds a plan from typed, domain-specific operators, validates it before execution, and reuses prior operator sequences for similar queries by letting a smaller LLM adapt only their parameters.

Across human activity recognition, ECG monitoring, and vehicular telematics datasets, Flash-Fusion reaches 92.2\% query accuracy, compared with 76.1\% for ReAct, the strongest baseline, while completing queries \(7\times\) faster and consuming \(5\times\) fewer tokens.
Schema grounding, typed plan validation, and reuse thus let users across disciplines explore IoT data without writing analysis code.

Flash-Fusion turns a query into a validated plan, but sensors never stop producing data. Users want to state a goal once and trust the system to watch for it.
We envision users expressing monitoring goals in natural language, with verifiable plans continuously evaluating incoming telemetry and raising alerts backed by evidence.
Two advances would make this practical.
First, schema grounding must extend to heterogeneous sources, which matters as sensor-language models widen the range of signals LLMs can interpret~\cite{li2025sensorllm,mo2024iot}.
Second, small local models must adapt reusable plans without cloud assistance, in line with early work placing LLMs on edge devices for IoT~\cite{belcak2025small}. Typed operators could then serve as a shared substrate through which agents request analyses and exchange results, a role that surveys of agentic data systems identify as an open need for reliable, interoperable agent workflows~\cite{ehtesham2025survey}.
Our future work will investigate these extensions toward continuous, distributed sensor analytics.

%% file: refs.bib
@article{houssein2024internet,
  title={Internet of things in smart cities: Comprehensive review, open issues and challenges},
  author={Houssein, Essam H and Othman, Mahmoud A and Mohamed, Waleed M and Younan, Mina},
  journal={IEEE Internet of Things Journal},
  year={2024},
  publisher={IEEE}
}

@article{zhang2026agentic,
  title={Agentic plan caching: Test-time memory for fast and cost-efficient LLM agents},
  author={Zhang, Qizheng and Wornow, Michael and Olukotun, Kunle},
  journal={Advances in Neural Information Processing Systems},
  volume={38},
  pages={103270--103296},
  year={2026}
}

@article{zong2025integrating,
  title={Integrating large language models with internet of things: applications},
  author={Zong, Mingyu and Hekmati, Arvin and Guastalla, Michael and Li, Yiyi and Krishnamachari, Bhaskar},
  journal={Discover Internet of Things},
  volume={5},
  number={1},
  pages={2},
  year={2025},
  publisher={Springer}
}

@article{li2024dawn,
  title={The dawn of natural language to sql: Are we fully ready?},
  author={Li, Boyan and Luo, Yuyu and Chai, Chengliang and Li, Guoliang and Tang, Nan},
  journal={arXiv preprint arXiv:2406.01265},
  year={2024}
}

@inproceedings{fei2024extending,
  title={Extending context window of large language models via semantic compression},
  author={Fei, Weizhi and Niu, Xueyan and Zhou, Pingyi and Hou, Lu and Bai, Bo and Deng, Lei and Han, Wei},
  booktitle={Findings of the Association for Computational Linguistics: ACL 2024},
  pages={5169--5181},
  year={2024}
}

@article{xu2025towards,
  title={Towards large reasoning models: A survey of reinforced reasoning with large language models},
  author={Xu, Fengli and Hao, Qianyue and Zong, Zefang and Wang, Jingwei and Zhang, Yunke and Wang, Jingyi and Lan, Xiaochong and Gong, Jiahui and Ouyang, Tianjian and Meng, Fanjin and others},
  journal={arXiv preprint arXiv:2501.09686},
  year={2025}
}

@article{liu2025comprehensive,
  title={A comprehensive survey on long context language modeling},
  author={Liu, Jiaheng and Zhu, Dawei and Bai, Zhiqi and He, Yancheng and Liao, Huanxuan and Que, Haoran and Wang, Zekun and Zhang, Chenchen and Zhang, Ge and Zhang, Jiebin and others},
  journal={arXiv preprint arXiv:2503.17407},
  year={2025}
}

@article{an2024iot,
  title={Iot-llm: Enhancing real-world iot task reasoning with large language models},
  author={An, Tuo and Zhou, Yunjiao and Zou, Han and Yang, Jianfei},
  journal={arXiv preprint arXiv:2410.02429},
  year={2024}
}

@article{al2025vega,
  title={Vega: LLM-Driven Intelligent Chatbot Platform for Internet of Things Control and Development},
  author={Al-Safi, Harith and Ibrahim, Harith and Steenson, Paul},
  journal={Sensors},
  volume={25},
  number={12},
  pages={3809},
  year={2025},
  publisher={MDPI}
}

@article{belcak2025small,
  title={Small language models are the future of agentic ai},
  author={Belcak, Peter and Heinrich, Greg and Diao, Shizhe and Fu, Yonggan and Dong, Xin and Muralidharan, Saurav and Lin, Yingyan Celine and Molchanov, Pavlo},
  journal={arXiv preprint arXiv:2506.02153},
  year={2025}
}

@article{ehtesham2025survey,
  title={A survey of agent interoperability protocols: Model context protocol (mcp), agent communication protocol (acp), agent-to-agent protocol (a2a), and agent network protocol (anp)},
  author={Ehtesham, Abul and Singh, Aditi and Gupta, Gaurav Kumar and Kumar, Saket},
  journal={arXiv preprint arXiv:2505.02279},
  year={2025}
}

@inproceedings{hadiyoso2015arrhythmia,
  title={Arrhythmia detection based on ECG signal using Android mobile for athlete and patient},
  author={Hadiyoso, Sugondo and Usman, Koredianto and Rizal, Achmad},
  booktitle={2015 3rd International Conference on Information and Communication Technology (ICoICT)},
  pages={166--171},
  year={2015},
  organization={IEEE}
}

@article{couderc2012telemetric,
  title={The Telemetric and Holter ECG Warehouse (THEW): the first three years of development and research},
  author={Couderc, Jean-Philippe},
  journal={Journal of electrocardiology},
  volume={45},
  number={6},
  pages={677--683},
  year={2012},
  publisher={Elsevier}
}

@article{colvin2024pydantic,
  title={Pydantic},
  author={Colvin, Samuel and Jolibois, Eric and Ramezani, Hasan and Garcia Badaracco, Adrian and Dorsey, Terrence and Montague, David and Matveenko, Serge and Trylesinski, Marcelo and Runkle, Sydney and Hewitt, David and others},
  journal={Zenodo},
  year={2024}
}

@online{groqpricing,
	title = {Groq On-demand Pricing for Tokens-as-a-Service},
    author={Ross, Jonathan},
	url = {https://groq.com/pricing},
	titleaddon = {Groq},
	urldate = {2025-11-05},
    year={2025},
	langid = {english},
}

@misc{openrouter_pricing,
  author = {{OpenRouter}},
  title  = {OpenRouter Pricing},
  year   = {2026},
  url    = {https://openrouter.ai/pricing}
}

@article{killeen2019iot,
  title={IoT-based predictive maintenance for fleet management},
  author={Killeen, Patrick and Ding, Bo and Kiringa, Iluju and Yeap, Tet},
  journal={Procedia Computer Science},
  volume={151},
  pages={607--613},
  year={2019},
  publisher={Elsevier}
}

@inproceedings{bang2023gptcache,
  title={Gptcache: An open-source semantic cache for llm applications enabling faster answers and cost savings},
  author={Bang, Fu},
  booktitle={Proceedings of the 3rd Workshop for Natural Language Processing Open Source Software (NLP-OSS 2023)},
  pages={212--218},
  year={2023}
}

@misc{ministralblog,
  title        = {Ministral 8B},
  author       = {{Mistral AI}},
  year         = {2024},
  month        = oct,
  url          = {https://docs.mistral.ai/models/ministral-8b-24-1},
  note         = {Accessed: 2026-10-22}
}

@misc{qwen3maxblog,
  title        = {Qwen3-Max: Just Scale it},
  author       = {{Qwen Team}},
  year         = {2025},
  month        = sep,
  url          = {https://qwen.ai/blog?id=qwen3-max},
  note         = {Accessed: 2026-10-22}
}

@article{mckinney2011pandas,
  title={pandas: a foundational Python library for data analysis and statistics},
  author={McKinney, Wes and others},
  journal={Python for high performance and scientific computing},
  volume={14},
  number={9},
  pages={1--9},
  year={2011},
  publisher={Seattle}
}

@article{mussah2023accelerating,
  title={Accelerating statewide connected vehicles big (sensor fusion) data ETL pipelines on GPUs},
  author={Mussah, Abdul Rashid and Shoman, Maged and Amo-Boateng, Mark and Adu-Gyamfi, Yaw},
  journal={arXiv preprint arXiv:2305.07454},
  year={2023}
}

@inproceedings{stabler2025impact,
  title={The impact of chunking strategies on domain-specific information retrieval in RAG systems},
  author={St{\"a}bler, Maximilian and Turnbull, Steffen and M{\"u}ller, Tobias and Langdon, Chris and Marx-Gom{\'e}z, Jorge and K{\"o}ster, Frank},
  booktitle={2025 IEEE International Conference on Omni-layer Intelligent Systems (COINS)},
  pages={1--6},
  year={2025},
  organization={IEEE}
}

@article{comanici2025gemini,
  title={Gemini 2.5: Pushing the frontier with advanced reasoning, multimodality, long context, and next generation agentic capabilities},
  author={Comanici, Gheorghe and Bieber, Eric and Schaekermann, Mike and Pasupat, Ice and Sachdeva, Noveen and Dhillon, Inderjit and Blistein, Marcel and Ram, Ori and Zhang, Dan and Rosen, Evan and others},
  journal={arXiv preprint arXiv:2507.06261},
  year={2025}
}

@inproceedings{shen2025autoiot,
  title={Autoiot: Llm-driven automated natural language programming for aiot applications},
  author={Shen, Leming and Yang, Qiang and Zheng, Yuanqing and Li, Mo},
  booktitle={Proceedings of the 31st Annual International Conference on Mobile Computing and Networking},
  pages={468--482},
  year={2025}
}

@article{ouyang2024llmsense,
  title={LLMSense: Harnessing LLMs for high-level reasoning over spatiotemporal sensor traces},
  author={Ouyang, Xiaomin and Srivastava, Mani},
  journal={arXiv preprint arXiv:2403.19857},
  year={2024}
}

@article{yao2022react,
  title={React: Synergizing reasoning and acting in language models},
  author={Yao, Shunyu and Zhao, Jeffrey and Yu, Dian and Du, Nan and Shafran, Izhak and Narasimhan, Karthik and Cao, Yuan},
  journal={arXiv preprint arXiv:2210.03629},
  year={2022}
}

@article{ye2021rapidfuzz,
  title={RapidFuzz: Accelerating fuzzing via generative adversarial networks},
  author={Ye, Aoshuang and Wang, Lina and Zhao, Lei and Ke, Jianpeng and Wang, Wenqi and Liu, Qinliang},
  journal={Neurocomputing},
  volume={460},
  pages={195--204},
  year={2021},
  publisher={Elsevier}
}

@inproceedings{yin2024study,
  title={A study of sentence similarity based on the all-minilm-l6-v2 model with “same semantics, different structure” after fine tuning},
  author={Yin, Chen and Zhang, Zixuan},
  booktitle={2024 2nd International Conference on Image, Algorithms and Artificial Intelligence (ICIAAI 2024)},
  pages={677--684},
  year={2024},
  organization={Atlantis Press}
}

@article{kalai2025language,
  title={Why language models hallucinate},
  author={Kalai, Adam Tauman and Nachum, Ofir and Vempala, Santosh S and Zhang, Edwin},
  journal={arXiv preprint arXiv:2509.04664},
  year={2025}
}

@article{roustan2025clinicians,
  title={The clinicians’ guide to large language models: a general perspective with a focus on hallucinations},
  author={Roustan, Dimitri and Bastardot, Fran{\c{c}}ois},
  journal={Interactive journal of medical research},
  volume={14},
  number={1},
  pages={e59823},
  year={2025},
  publisher={JMIR Publications Inc., Toronto, Canada}
}

@article{wei2022chain,
  title={Chain-of-thought prompting elicits reasoning in large language models},
  author={Wei, Jason and Wang, Xuezhi and Schuurmans, Dale and Bosma, Maarten and Xia, Fei and Chi, Ed and Le, Quoc V and Zhou, Denny and others},
  journal={Advances in neural information processing systems},
  volume={35},
  pages={24824--24837},
  year={2022}
}

@inproceedings{ji2024hargpt,
  title={Hargpt: Are llms zero-shot human activity recognizers?},
  author={Ji, Sijie and Zheng, Xinzhe and Wu, Chenshu},
  booktitle={2024 IEEE International Workshop on Foundation Models for Cyber-Physical Systems \& Internet of Things (FMSys)},
  pages={38--43},
  year={2024},
  organization={IEEE}
}

@article{weiss2019wisdm,
  title={Wisdm smartphone and smartwatch activity and biometrics dataset},
  author={Weiss, Gary M},
  journal={UCI Machine Learning Repository: WISDM Smartphone and Smartwatch Activity and Biometrics Dataset Data Set},
  volume={7},
  number={133190-133202},
  pages={5},
  year={2019},
  publisher={Springer}
}

@inproceedings{apandi2018arrhythmia,
  title={Arrhythmia detection using MIT-BIH dataset: A review},
  author={Apandi, Ziti Fariha Mohd and Ikeura, Ryojun and Hayakawa, Soichiro},
  booktitle={2018 International Conference on Computational Approach in Smart Systems Design and Applications (ICASSDA)},
  pages={1--5},
  year={2018},
  organization={IEEE}
}

@inproceedings{wu2025tablebench,
  title={Tablebench: A comprehensive and complex benchmark for table question answering},
  author={Wu, Xianjie and Yang, Jian and Chai, Linzheng and Zhang, Ge and Liu, Jiaheng and Du, Xeron and Liang, Di and Shu, Daixin and Cheng, Xianfu and Sun, Tianzhen and others},
  booktitle={Proceedings of the AAAI Conference on Artificial Intelligence},
  volume={39},
  number={24},
  pages={25497--25506},
  year={2025}
}

@article{li2023can,
  title={Can llm already serve as a database interface? a big bench for large-scale database grounded text-to-sqls},
  author={Li, Jinyang and Hui, Binyuan and Qu, Ge and Yang, Jiaxi and Li, Binhua and Li, Bowen and Wang, Bailin and Qin, Bowen and Geng, Ruiying and Huo, Nan and others},
  journal={Advances in Neural Information Processing Systems},
  volume={36},
  pages={42330--42357},
  year={2023}
}

@inproceedings{rajpurkar2018know,
  title={Know what you don’t know: Unanswerable questions for SQuAD},
  author={Rajpurkar, Pranav and Jia, Robin and Liang, Percy},
  booktitle={Proceedings of the 56th Annual Meeting of the Association for Computational Linguistics (Volume 2: Short Papers)},
  pages={784--789},
  year={2018}
}

@article{zheng2023judging,
  title={Judging llm-as-a-judge with mt-bench and chatbot arena},
  author={Zheng, Lianmin and Chiang, Wei-Lin and Sheng, Ying and Zhuang, Siyuan and Wu, Zhanghao and Zhuang, Yonghao and Lin, Zi and Li, Zhuohan and Li, Dacheng and Xing, Eric and others},
  journal={Advances in neural information processing systems},
  volume={36},
  pages={46595--46623},
  year={2023}
}

@online{gpt-55,
	title = {{GPT}-5.5 Model {\textbar} {OpenAI} {API}},
  author = {OpenAI},
	url = {https://developers.openai.com/api/docs/models/gpt-5.5},
	urldate = {2026-06-04},
	langid = {english},
  year = {2026}
}

@article{patel2025semantic,
  title={Semantic operators and their optimization: Enabling llm-based data processing with accuracy guarantees in lotus},
  author={Patel, Liana and Jha, Siddharth and Pan, Melissa and Gupta, Harshit and Asawa, Parth and Guestrin, Carlos and Zaharia, Matei},
  journal={Proceedings of the VLDB Endowment},
  volume={18},
  number={11},
  pages={4171--4184},
  year={2025},
  publisher={VLDB Endowment}
}

@inproceedings{liu2025palimpzest,
  title={Palimpzest: Optimizing AI-Powered Analytics with Declarative Query Processing.},
  author={Liu, Chunwei and Russo, Matthew and Cafarella, Michael J and Cao, Lei and Chen, Peter Baile and Chen, Zui and Franklin, Michael J and Kraska, Tim and Madden, Samuel and Shahout, Rana and others},
  booktitle={CIDR},
  year={2025}
}

@article{russo2025abacus,
  title={Abacus: A cost-based optimizer for semantic operator systems},
  author={Russo, Matthew and Liu, Chunwei and Sudhir, Sivaprasad and Vitagliano, Gerardo and Cafarella, Michael and Kraska, Tim and Madden, Samuel},
  journal={arXiv preprint arXiv:2505.14661},
  year={2025}
}

@article{lao2025sembench,
  title={Sembench: A benchmark for semantic query processing engines},
  author={Lao, Jiale and Zimmerer, Andreas and Ovcharenko, Olga and Cong, Tianji and Russo, Matthew and Vitagliano, Gerardo and Cochez, Michael and {\"O}zcan, Fatma and Gupta, Gautam and Hottelier, Thibaud and others},
  journal={arXiv preprint arXiv:2511.01716},
  year={2025}
}

@article{zhang2026sensorlm,
  title={SensorLM: Learning the language of wearable sensors},
  author={Zhang, Yuwei and Ayush, Kumar and Qiao, Siyuan and Heydari, A Ali and Narayanswamy, Girish and Xu, Max and Metwally, Ahmed and Xu, Jinhua and Garrison, Jake and Xu, Xuhai and others},
  journal={Advances in Neural Information Processing Systems},
  volume={38},
  pages={46813--46849},
  year={2026}
}

@inproceedings{li2025sensorllm,
  title={Sensorllm: Aligning large language models with motion sensors for human activity recognition},
  author={Li, Zechen and Deldari, Shohreh and Chen, Linyao and Xue, Hao and Salim, Flora D},
  booktitle={Proceedings of the 2025 Conference on Empirical Methods in Natural Language Processing},
  pages={354--379},
  year={2025}
}

@article{mo2024iot,
  title={Iot-lm: Large multisensory language models for the internet of things},
  author={Mo, Shentong and Salakhutdinov, Russ and Morency, Louis-Philippe and Liang, Paul Pu},
  journal={arXiv preprint arXiv:2407.09801},
  year={2024}
}

@inproceedings{asif2025llasa,
  title={LLaSA: A Sensor-Aware LLM for Natural Language Reasoning of Human Activity from IMU Data},
  author={Asif Imran Shouborno, Sheikh and Khan, Mohammad Nur Hossain and Biswas, Subrata and Islam, Bashima},
  booktitle={Companion of the 2025 ACM International Joint Conference on Pervasive and Ubiquitous Computing},
  pages={893--899},
  year={2025}
}

@article{zhou2026holollm,
  title={Holollm: Multisensory foundation model for language-grounded human sensing and reasoning},
  author={Zhou, Chuhao and Yang, Jianfei},
  journal={Advances in Neural Information Processing Systems},
  volume={38},
  pages={73543--73568},
  year={2026}
}

@article{wang2024agent,
  title={Agent workflow memory},
  author={Wang, Zora Zhiruo and Mao, Jiayuan and Fried, Daniel and Neubig, Graham},
  journal={arXiv preprint arXiv:2409.07429},
  year={2024}
}

@inproceedings{pham2026agentstop,
  title={AgentStop: Terminating Local AI Agents Early to Save Energy in Consumer Devices},
  author={Pham, Dzung and Katevas, Kleomenis and Shahin Shamsabadi, Ali and Haddadi, Hamed},
  booktitle={Proceedings of the ACM Conference on AI and Agentic Systems},
  pages={1051--1069},
  year={2026}
}

@online{langchain_errors_2026,
  title        = {Tool Integrations},
  author       = {{LangChain}},
  organization = {LangChain},
  url          = {https://docs.langchain.com/oss/python/integrations/tools},
  titleaddon   = {LangChain Documentation},
  date      = {2026-06-23},
  langid       = {english},
}

@online{huggingface,
  title   = {Hugging Face Hub},
  author  = {{Hugging Face}},
  url     = {https://huggingface.co/},
  date = {2026-10-22},
  langid  = {english},
}

@online{snowball,
  title   = {Snowball},
  author  = {{Snowball Project}},
  url     = {https://snowballstem.org/},
  date = {2026-11-04},
  langid  = {english},
}
